# Subnanometer localization accuracy in widefield optical microscopy

Craig R. Copeland[1,2], Jon Geist[3], Craig D. McGray[3], Vladimir A. Aksyuk[1], J. Alexander Liddle[1], B. Robert Ilic[1], and Samuel M. Stavis[1,*]

[1]Center for Nanoscale Science and Technology, National Institute of Standards and Technology, Gaithersburg, Maryland 20899, United States of America, [2]Maryland NanoCenter, University of Maryland, College Park, Maryland 20742, United States of America, [3]Engineering Physics Division, National Institute of Standards and Technology, Gaithersburg, Maryland 20899, United States of America

**Abstract:** The common assumption that precision is the limit of accuracy in localization microscopy and the typical absence of comprehensive calibration of optical microscopes lead to a widespread issue – overconfidence in measurement results with nanoscale statistical uncertainties that can be invalid due to microscale systematic errors. In this article, we report a comprehensive solution to this underappreciated problem. We develop arrays of subresolution apertures into the first reference materials that enable localization errors approaching the atomic scale across a submillimeter field. We present novel methods for calibrating our microscope system using aperture arrays and develop aberration corrections that reach the precision limit of our reference materials. We correct and register localization data from multiple colors and test different sources of light emission with equal accuracy, indicating the general applicability of our reference materials and calibration methods. In a first application of our new measurement capability, we introduce the concept of critical dimension localization microscopy, facilitating tests of nanofabrication processes and quality control of aperture arrays. In a second application, we apply these stable reference materials to answer open questions about the apparent instability of fluorescent nanoparticles that commonly serve as fiducial markers. Our study establishes a foundation for subnanometer localization accuracy in widefield optical microscopy.

**Keywords:** critical dimension; localization accuracy; nanoparticle fiducial; optical microscopy; reference material

## INTRODUCTION

Optical microscopy methods of localizing small emitters are broadly useful in such fields as cell biology, nanoscale fabrication, cryogenic physics, and microelectromechanical systems[1]. Both precision[2,3,4] and accuracy are fundamental to localization microscopy[5,6]. Localization of single fluorophores with a statistical uncertainty of tens of nanometers is common, and subnanometer uncertainty is possible for fluorophores[7] and readily achievable for brighter emitters such as particles[8]. Whereas improving localization precision generally requires counting more signal photons by increasing the intensity and stability of emission[9,10], achieving commensurate localization accuracy presents diverse challenges in the calibration of an optical microscope as a nonideal measurement system. Such calibration involves not only the discrete parts of the system but also the interaction of those parts during a measurement and is rarely, if ever, implemented. This can cause overconfidence in measurement results with statistical uncertainties at the nanometer scale that are invalid due to larger systematic errors. These errors can extend into the micrometer scale when localizing emitters across a wide field, as is often necessary for imaging microstructures and tracking motion[11,12]. The discrepancy between precision and accuracy can be so large as to require a logarithmic target to illustrate, as Figure 1 shows.

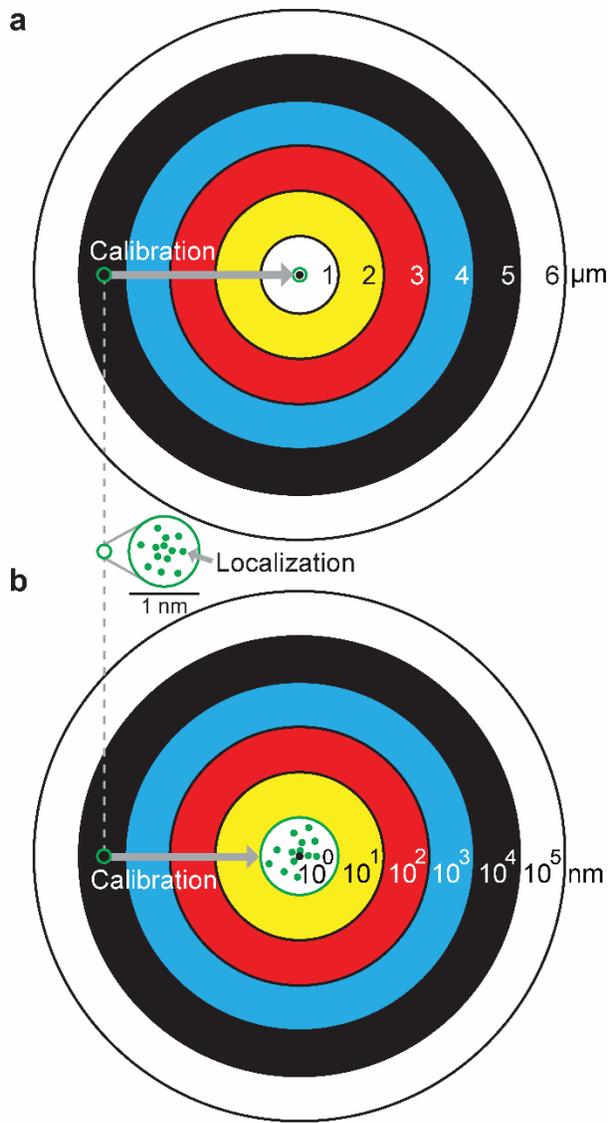

**Figure 1.** (**a**) Schematic showing a linear target. (**b**) Schematic showing a logarithmic target. Green dots are localization data. Their scatter indicates statistical uncertainty at the subnanometer scale, which is not apparent on the linear target as systematic errors can be four orders of magnitude larger. This discrepancy requires a logarithmic target to illustrate both precision and accuracy. Calibration of the measurement system and correction of localization data ensures that precision is the limit of accuracy[13].

The root cause of the problem is a lack of reference materials and calibration methods that are optimal for localization microscopy, analogous to those for optical imaging at larger scales[14]. Small particles are useful for mapping certain effects of optical aberrations[15, 16, 17]. However, their size distribution and random deposition can result in nonuniform sampling of the imaging field, fluorophores in particles often have a different emission spectrum from that of fluorophores in solution, and evaluating magnification[18] requires a specification of distance between emitters. DNA origami can control the submicrometer distance between a few fluorophores[19, 20], but this approach has limitations of emitter intensity and stability, as well as sampling uniformity. Stages require their own calibration to scan emitters through the imaging field, while microscope instability can limit sampling accuracy[21, 22, 23]. Arrays of subresolution apertures enable calibration of both aberrations and magnification, with intense and stable emission, and uniform and accurate sampling[24]. Recent studies have used aperture arrays to calibrate the effects of chromatic aberrations on image registration[22, 23, 25, 26], sample orientation and aberrations in three dimensions[27] and image pixel size[28]. However, these studies have not quantified the critical dimensions of an aperture array to produce a reference material, demonstrated all functions of an aperture array for microscope calibration, or reached the performance limits of the corresponding calibration methods. Other factors contribute to the overall problem, as follows.

Electron-multiplying charge-coupled-device (EMCCD) cameras were common at the advent of localization microscopy and their calibration continues[29]. Complementary metal-oxide-semiconductor (CMOS) cameras are of increasing interest due to advantages of performance and cost but have nonuniform sensitivity and read noise. Initial studies tested the effects of CMOS noise on localization[30] and improved the localization of single fluorophores[31, 32]. However, no study has calibrated over the full dynamic range of a CMOS camera to maximize the number of signal photons and minimize statistical uncertainty. Previous studies have improved illumination uniformity[33] and performed flatfield corrections but have not accounted for all related CMOS nonuniformities.

Localization analysis extracts information from optical images. Maximum-likelihood and weighted least-squares algorithms[34, 35], with specific estimators for CMOS cameras[31, 32], compete on the basis of accuracy and efficiency. However, previous studies have not evaluated the performance of each algorithm in the presence of discrepancies between model approximations of the point spread function and experimental data. The resulting fitting errors are common for models that neglect deformations from aberrations[36, 37, 38], which vary across a wide field.

Finally, localization of a fiducial marker such as a small particle often provides a reference position for correcting systematic errors from unintentional motion of the sample or microscope[9, 39, 40, 41]. A typical but critical assumption is that the fiducial is motionless with respect to the sample. However, there are open questions about whether nanoparticle fiducials are truly static on imaging substrates[15, 34, 39]. Confounding this issue, microscope systems are not perfectly stable, and there is no appropriate reference material for assessing their subnanometer stability across a wide field.

In this study, we present a comprehensive solution to this overall problem, reducing localization errors from a widefield optical microscope by up to four orders of magnitude and transforming the microscope into a quantitative metrology system. We develop aperture arrays into prototype reference materials with multiple functions and combine them with novel methods to calibrate the parts of the system and their interaction during a measurement. We validate our widefield measurements and quantify localization error approaching the scale of atomic diameters across a submillimeter field, for multiple colors and emission sources. We apply our new measurement capability to introduce the concept of critical dimension localization microscopy of aperture arrays and to answer open questions about the apparent motion of nanoparticle fiducials. By minimizing and quantifying systematic errors at subnanometer scales, we enable rigorous confidence in precision as the limit of accuracy for localization microscopy.

## MATERIALS AND METHODS

**Aperture arrays**

We design[42] and fabricate square arrays of circular apertures with nominal diameters ranging from 200 nm to 500 nm in titanium and platinum films with a total thickness of approximately 100 nm on silica substrates with a thickness of approximately 170 µm. We use two different electron-beam lithography systems to pattern independent arrays and test the accuracy of aperture placement. Both lithography systems have traceable laser interferometers that measure stage position with a resolution of approximately 0.6 nm in the x and y directions to calibrate beam position and to confirm the absence of, or correct for, electron-optical aberrations. To avoid additional errors of aperture placement from stage motion of the lithography systems, we limit the lateral extents of our arrays to single write fields. Further details are in Supplementary Notes S1 and S2, Supplementary Table S1, and Supplementary Figs. S1-S4. To develop our calibration methods, we initially assume placement accuracy and we assume that random errors determine placement precision, as we define in Supplementary Table S2. We subsequently measure these dimensional properties.

**Fluorescent samples**

For some measurements, we fill the aperture array with a solution of boron-dipyrromethene dye at a concentration of approximately 200 µM in N,N-dimethylformamide. We also test fluorescent nanoparticles as fiducial markers. The manufacturer specifies polystyrene spheres with a mean diameter of 220 nm, containing boron-dipyrromethene dye molecules and a carboxylic acid coating. We disperse the nanoparticles in pure water, deposit 10 µL of the suspension onto a borosilicate coverslip with a thickness of approximately 170 µm and a poly-D-lysine coating, and remove the suspension after 1 min. We expect the nanoparticles to bind electrostatically to the coverslip. We cover the sample surface with pure water and seal it with another borosilicate coverslip for imaging. The emission spectra of the fluorescent dyes in solution and in nanoparticles are in Supplementary Fig. S6.

**Optical microscope**

Our microscope has an inverted stand, a scanning stage that translates in the x and y directions with a sample holder that rotates around these axes, and a piezoelectric actuator that translates an objective lens in the z direction with a nominal resolution of 10 nm. We typically use an objective lens with a nominal magnification of 63×, a numerical aperture of 1.2, and an immersion medium with an index of refraction of 1.33, resulting in a nominal depth of field of 0.95 µm at a wavelength of 500 nm. We reconfigure the microscope to epiilluminate fluorescent dye in aperture arrays and fluorescent nanoparticles on a microscope coverslip or transilluminate empty aperture arrays with a light-emitting diode (LED) array. The numerical aperture of the transilluminator condenser is 0.55. The emission spectra for the three LED arrays that we use are in Supplementary Fig. S6. The microscope has a CMOS camera with 2048 pixels by 2048 pixels, each with an on-chip size of 6.5 µm by 6.5 µm. We always operate the camera with water cooling and without on-board correction of pixel noise. We typically operate the camera in fast-scan mode, cool the sensor to -10 °C, and calibrate the imaging system for these parameters. In tests of fiducial stability, we operate the camera in slow-scan mode and cool the sensor to -30 °C. For fluorescence imaging, we use an excitation filter with a bandwidth from 450 nm to 500 nm, a dichroic mirror with a transition at 505 nm, and an emission filter with a bandwidth from 515 nm to 565 nm. We always equilibrate the microscope for at least 1 h before acquiring data. Representative micrographs of an aperture array and nanoparticle fiducials are in Supplementary Figs. S3-S5.

**Sample orientation and position**

We level the aperture array by iteratively rotating it around its x and y axes, and translating the objective lens in the z direction to simultaneously focus on apertures at the four corners of the imaging field. We test an alternate method for leveling the sample by analysis of Zernike coefficients, as Supplementary Note S3 describes. A schematic of our sample holder and corresponding results are in Supplementary Fig. S7. For all measurements, unless we note otherwise, we translate the objective lens through z to obtain a series of images around optimal focus for each aperture in an array, as Supplementary Note S4 describes and Supplementary Fig. S8 shows. We image at array centers unless we note otherwise.

**Camera calibration**

For each pixel $i$, we measure pixel value offset $o_i$ as the mean and read noise $\sigma^2_{\text{read},i}$ as the variance of 60 000 images[31] with the camera shutter closed. We determine flatfield corrections by imaging a white, planar object that is far out of focus and effectively featureless, at nine illumination levels spanning the dynamic range of the imaging sensor, $FF_i = \frac{\overline{I_i^*} - o_i}{\overline{I}}$, where $\overline{I_i}$ is the mean value of pixel $i$ from 15 000 images at an illumination level, $o_i$ is the pixel value offset, and $\overline{I}$ is the mean value of $\overline{I_i^*} - o_i$ from all pixels. The total noise of each pixel is the variance of the pixel value minus the pixel value offset from the 15 000 images at each illumination level. Plots and histograms of pixel value offset and read noise are in Supplementary Fig. S9.

**Model fitting**

We fit polynomial models to data using unweighted least-squares estimation and the Levenberg–Marquardt algorithm to determine optimal focus, characterize CMOS response, and calculate Zernike coefficients. We fit Gaussian models to images of point spread functions using various estimators and the Nelder-Mead simplex algorithm[43] to localize single emitters.

**RESULTS AND DISCUSSION**

**Terminology**

For processes ranging from aperture fabrication to data registration, we define qualitative terms, sources of error, and corresponding quantities in Supplementary Table S2. Our terminology is consistent with both common use and a common guide for metrology vocabulary[13].

**Aperture array**

We test epiillumination of a fluorescent dye in the apertures[27] and transillumination of empty apertures[23] as relevant configurations for localization microscopy. Whereas the dye solution degrades and requires cleaning, empty apertures are more stable and thus appropriate for developing our calibration methods. After doing so, we revisit the difference between the two configurations. Transillumination of empty apertures produces an array of point sources, as Figure 2 and Supplementary Fig. S3 show, and as Supplementary Table S1 and Supplementary Note S7 describe. An array pitch of at least 5 µm ensures that the point spread functions of adjacent apertures do not overlap significantly, as Supplementary Fig. S4 shows.

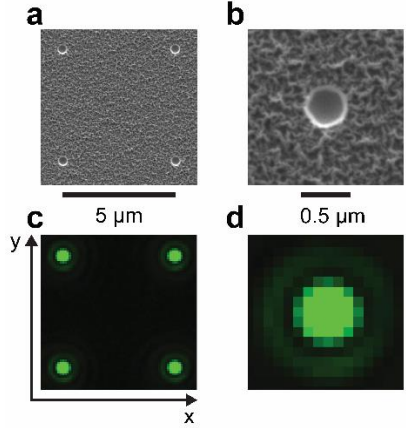

**Figure 2.** Aperture array. **(a-b)** Scanning electron micrographs showing representative apertures in a metal bilayer on a silica substrate. (a) The array has a nominal pitch of 5 µm. (b) Apertures have nominal diameters of 400 nm and smaller functional diameters, as Supplementary Note S7 and Supplementary Table S1 describe. **(c-d)** Brightfield optical micrographs showing representative apertures transmitting light. False color represents the peak illumination wavelength of 500 nm. (c) Four apertures form unit cells for pitch analysis. (d) The image of an aperture closely resembles the point spread function of the imaging system.

**CMOS calibration**

Accurate localization of aperture images first requires calibration of our CMOS camera, which we find is even less uniform than indicated by previous studies. Nonuniform pixel gain, sensor packaging, and illumination intensity cause significant variation in pixel value, motivating a flatfield correction. This correction increases with pixel value mean through the bottom 5 % of the dynamic range and then remains nearly constant over the remaining 95 %, as Supplementary Fig. S10a-b shows. A recent study did not identify this trend but presented localization algorithms that still achieved the Cramér–Rao lower bound[32]. Therefore, we use the constant correction in our analysis of pixel values that span the full dynamic range. Total noise, or pixel value variance, including read noise, shot noise, and fixed-pattern noise, does not depend linearly on pixel value mean over the full dynamic range, as Supplementary Fig. S10c-d shows, in contrast to a linear approximation from Poisson statistics at low pixel values. A quartic polynomial is a better approximation, but the linear approximation results in localization that is equally accurate and more efficient. Further details are in Supplementary Note S5 and Supplementary Table S3.

**Localization algorithm**

Aberrations, such as from objective lenses[44], can become significant across a wide field and deform the point spread function in ways that are typically unpredictable. Most localization algorithms do not account for such deformation, and one even requires its absence[45]. Previous studies have not fully explored the effects of fitting errors[35, 46] on the performance of weighted least-squares[32] or maximum-likelihood[31] estimation. These algorithms can include information from CMOS calibration and shot noise, unlike unweighted least-squares. There are arguments for and against each algorithm[32, 34]. Rather than strictly adhering to one algorithm or another, we use the aperture array to test their performance in the presence of fitting errors from aberration effects, which vary across a wide field. For this test, we select a bivariate Gaussian approximation of the point spread function,

$$G_{biv}(x,y) = A \cdot \exp - \left( \frac{1}{2(1-\rho^2)} \left[ \frac{(x-x_0)^2}{\sigma_x^2} - 2\rho \frac{(x-x_0)(y-y_0)}{\sigma_x \sigma_y} + \frac{(y-y_0)^2}{\sigma_y^2} \right] \right) + C, \quad (\text{Eq. 1})$$

where $A$ is the amplitude, $x_0$ is the position of the peak in the x direction, $y_0$ is the position of the peak in the y direction, $\sigma_x$ is the standard deviation in the x direction, $\sigma_y$ is the standard deviation in the y direction, $\rho$ is the correlation coefficient between the x and y directions, and $C$ is a constant background. Unlike a univariate Gaussian function, this model has some empirical ability to accommodate asymmetry from deformation of the point spread function[24, 47], which can be significant, as Figure 3 shows at a corner of the imaging field, 140 µm away from its center.

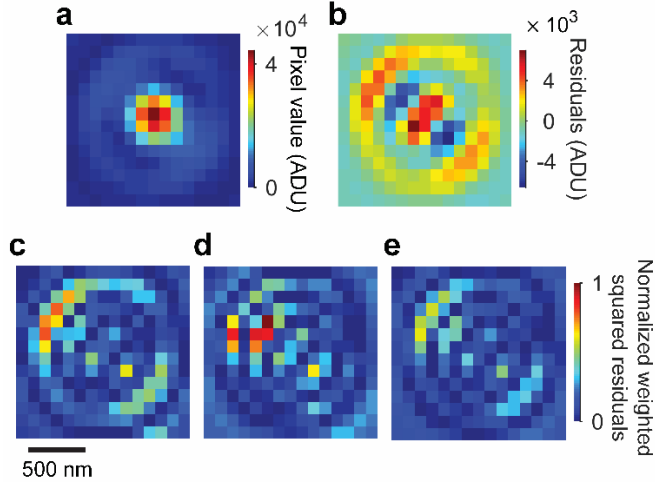

**Figure 3.** Localization algorithms. (**a**) Brightfield optical micrograph showing the localization region of interest containing a point spread function with asymmetry from aberrations. Pixel values are in analog-to-digital units (ADU). False color enhances contrast. We fit a bivariate Gaussian model to the data to test the estimation performance of three localization algorithms in the presence of model discrepancy. (**b**) Plot showing residuals from a fit using the light-weighting objective function. (**c-e**) Plots showing weighted squared residuals on a normalized scale. (c) Weighted least-squares heavily weights the first Airy ring. (d) Maximum-likelihood heavily weights between the central peak and Airy ring. (e) Light-weighting results in more uniform weighting than either (c) or (d) and improves empirical localization precision.

In light of the fitting errors that result, we introduce an empirical objective function for robust parameter estimation. The light-weighting objective function reduces the effect of fitting errors whether the model overestimates or underestimates the data,

$$\hat{\Theta} = \text{argmin} \left[ \sum_i \frac{(I_i - E_i)^2}{g \cdot \max(I_i, E_i) + \sigma_{\text{read},i}^2} \right], \quad (\text{Eq. 2})$$

where $\hat{\Theta}$ is the estimate for the parameter set $\hat{\Theta} = \{A, \sigma_x, \sigma_y, \rho, x_0, y_0, C\}$, $i$ indexes each pixel, $I_i$ is the experimental pixel value after CMOS calibration, $E_i$ is the expected or model pixel value, $g$ is the nominal gain of the camera, and $\sigma_{\text{read},i}^2$ is the pixel read noise. The use of $\max(I_i, E_i)$ selects either weighted least-squares ($I_i > E_i$) or maximum-likelihood ($I_i < E_i$) to reduce the weights of pixels with large residuals due to model discrepancy. Further details are in Supplementary Note S6.

The algorithm performance depends on both the deformation extent and the photon count, as Supplementary Fig. S11 and Supplementary Table S4 show. For our wide field and intense emitters, light-weighting improves empirical localization precision on average, as Supplementary

Table S4 shows. In field regions with large deformation, unweighted least-squares improves localization precision relative to the other algorithms. In field regions with small deformation, light-weighting, maximum-likelihood, and weighted least-squares perform comparably. The same is true when the localization region of interest excludes regions of the point spread function that cause the largest fitting errors, but doing so degrades empirical localization precision on average, as Supplementary Table S4 shows. We subsequently quantify localization error, including any effects of fitting errors.

**Aberration effects**

Aberrations degrade localization accuracy through several effects. In our experimental system, a silica substrate of standard thickness and high quality underpins the aperture array and is therefore part of the microscope system and its calibration. Additional calibration may be necessary for aberration effects from an experimental sample[48]. We begin to calibrate aberration effects by characterizing the bivariate Gaussian approximation of the point spread function in three dimensions. We image the aperture array through focus, and locate optimal focus for each aperture as the z position that maximizes the amplitude of the resulting point spread function, as Supplementary Fig. S8 shows. The field curves in the z direction over a range of nearly 500 nm, as Figure 4a,b show. We confirm the effective flatness of the aperture array, as Supplementary Fig. S2 shows. Without such characterization, a nonplanar array can corrupt calibration for localization in three dimensions[27]. The complex curvature of the field motivates the use of an aperture array to uniformly sample it, and has several consequences. Not all objects across the field can be at optimal focus simultaneously. Many experiments permit acquisition of only a single micrograph, which can be at a z position that maximizes the mean amplitude of point spread functions across the field. We define this optimal focal plane as z = 0 in Figure 4b. If the quasistatic imaging of stable emitters is feasible, then acquiring multiple micrographs along the curving field allows for optimal focus of each point spread function.

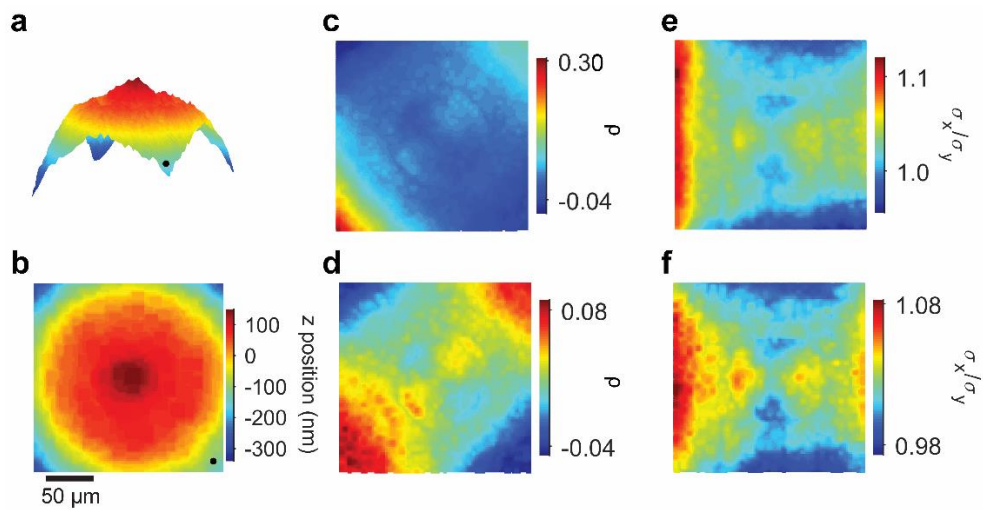

**Figure 4.** Field curvature and point spread function deformation. (**a, b**) Plots showing the curving field of the imaging system. The black dots mark the same corner. The optimal focal plane is at z=0 nm. (**c**) Plot showing a larger range of $\rho$ from a single image at the optimal focal plane, maximizing the mean amplitude of all point spread functions. (**d**) Plot showing a smaller range of $\rho$ from multiple images along the curving field, maximizing the amplitude of each point spread

function. (**e**) Plot showing $\sigma_x/\sigma_y$ from a single image at the optimal focal plane. (**f**) Plot showing $\sigma_x/\sigma_y$ from multiple images along the curving field. For these plots and subsequent plots showing optical effects, we use linear interpolations of data between aperture positions.

For the bivariate Gaussian approximation of the point spread function, the dimensionless parameters $\rho$ and $\sigma_x/\sigma_y$ describe asymmetries resulting from deformation. We extract these parameters from one image at the optimal focal plane, as Figure 4c,e show, and from multiple images along the curving field at which all apertures are in optimal focus, as Figure 4d,f show. In either case, the parameters have a similar field dependence. Imaging through focus reduces the range of $\rho$ by a factor of approximately three but has little effect on $\sigma_x/\sigma_y$. Either analysis can improve localization by fixing or improving initial guesses of model parameters in minimization algorithms, which can be important for localization accuracy.[35] These results also imply the potential for parameterizing more complex models of the point spread function, as well as for exploiting aberrations to localize emitters in three dimensions.

From one micrograph at the optimal focal plane, we localize each aperture and perform a similarity transformation to map an ideal array, with a pitch that is identical to the nominal value of 5 µm, to the localization data. This transformation consists of planar translation and rotation, and uniform scaling to determine the mean value of image pixel size. The differences between the positions that we measure and the nominal positions in the ideal array define position errors. The transformation scale factor results in a mean value of image pixel size of 99.94 nm, which is 3 % smaller than the nominal value of 103 nm. We revisit the uncertainty of image pixel size. Using the nominal value of image pixel size, which is a common but inadvisable practice, results in position errors of up to 4.5 µm, as Figure 5a-c show. Using the mean value of image pixel size resulting from the similarity transformation reduces these position errors by a factor of more than 18, however, the errors are still as large as 250 nm and vary nonmonotonically across the field, as Figure 5d-f show. These position errors are due primarily to pincushion distortion but also to field curvature and deformation of the point spread function. This extent of magnification calibration is comparable to that of a previous study that averaged over these effects in determining a mean value of image pixel size[18], and demonstrates the utility of sampling the field with an aperture array to further reduce systematic errors from aberration effects.

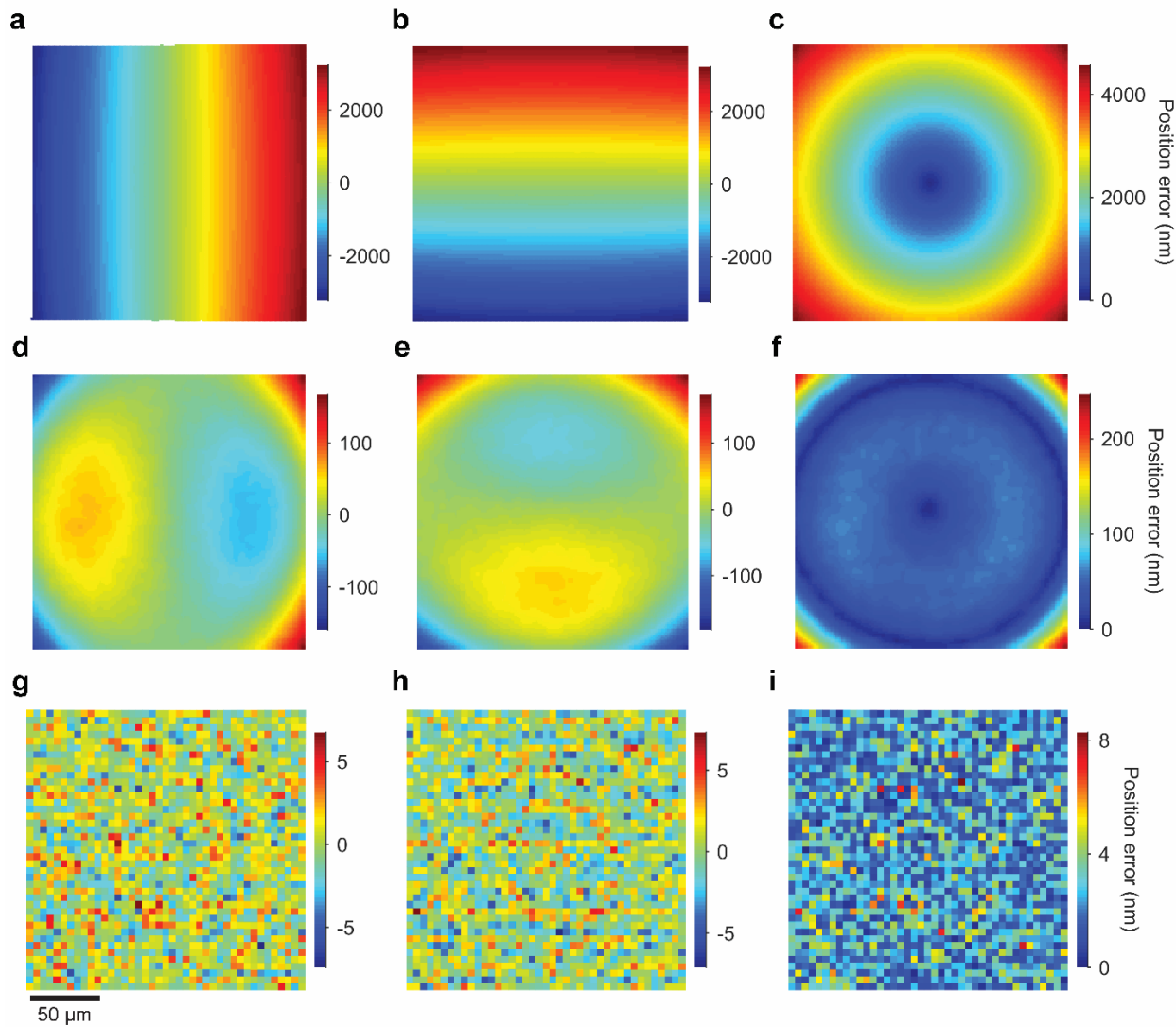

**Figure 5.** Position errors. **(a-c)** Plots showing position errors in (a) the x direction, (b) the y direction, and (c) total magnitude, due mostly to using the nominal value of image pixel size of 103 nm. **(d-f)** Plots showing position errors in (d) the x direction, (e) the y direction, and (f) total magnitude, after applying a similarity transformation to map the data in (a-c) to an ideal array, due mostly to using the mean value of image pixel size of 99.94 nm. **(g-i)** Plots showing position errors in (g) the x direction, (h) the y direction, and (i) total magnitude, after applying a correction model to the localization data in (d-f), due mostly to placement precision.

With other objective lenses, our microscope system shows comparable aberration effects of variable magnitude and field dependence, as Supplementary Fig. S12 and Supplementary Table S5 show. All of the objective lenses that we test result in mean values of image pixel size that are smaller than the nominal values by approximately 3 %, indicating that our microscope tube lens is the primary source of this systematic error. This finding is consistent with our observations of other microscope systems from the same manufacturer, which we do not show. The lens with the lowest numerical aperture results in the smallest position errors, revealing an unnecessary competition between collection efficiency and magnification uniformity that exists in the absence of calibration.

**Error correction**

We model the position errors in Figure 5d-f by a linear combination of consecutive Zernike polynomials[49] to develop a widefield correction that is applicable to position data from many forms of localization microscopy. The correction takes as input the inaccurate position of an emitter from a localization measurement, and gives as output its accurate position. The similarity transformation gives the value of image pixel size. At the center of the standard array from which we derive the model, the standard deviation of position error decreases monotonically with maximum Noll order, as Figure 6a shows. Sharp decreases correspond to polynomials with odd radial degrees greater than 1 and azimuthal degrees of 1 and -1, providing insight for optimization of the model by selection of a subset of nonconsecutive Zernike polynomials.

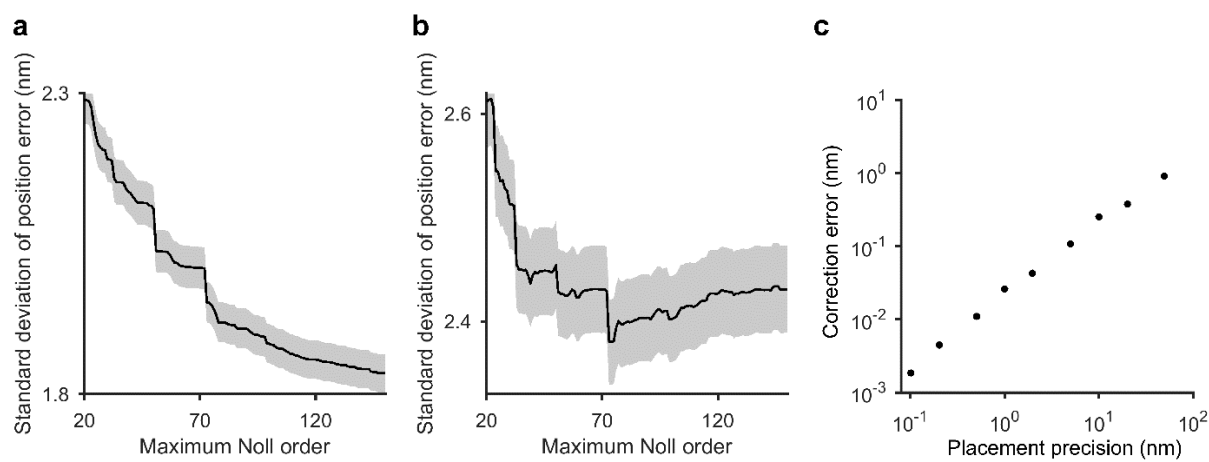

**Figure 6.** Correction model. (**a, b**) Plots showing representative values of the standard deviation of position errors in a single lateral dimension after correction, as a function of the number of consecutive Zernike polynomials in the model, or the maximum Noll order. A maximum Noll order of less than 20 corrects the largest fraction of the position errors. (a) At the center of the standard array from which we derive the model, the standard deviation decreases monotonically with maximum Noll order as the model corrects position errors due primarily to aberrations. (b) After applying the model from (a) to a different region of the standard array, the standard deviation decreases to a minimum at a maximum Noll order of 73 and then increases with additional orders, indicating erroneous inclusion of position errors due to placement precision at the array center. Plots for other regions of the array are similar. Gray bounds are one standard error. (**c**) Plot showing correction error, which increases approximately linearly with placement precision. Standard errors are smaller than the data markers.

We quantify the effect of placement precision on the correction model by two novel tests. First, we apply the correction to a different region of the standard array. The standard deviation of position error decreases to a minimum at a maximum Noll order of 73 and then increases, as Figure 6b shows. This trend indicates a limit beyond which additional consecutive Zernike polynomials erroneously correct position errors due to placement precision at the array center, degrading correction accuracy. To test this effect in the correction model of maximum Noll order 73, we simulate position errors due to placement precision as the standard deviation of a normal distribution around a mean pitch of 5 µm, and apply the correction to the resulting positions. The correction error depends approximately linearly on the magnitude of placement precision, as Figure 6c shows, and contributes less than 0.05 nm to the localization error for our aperture array.

The correction model of maximum Noll order 73 reduces the position errors in Figure 5d-f by another factor of 30, resulting in position errors in the x and y directions that are apparently random, as Figure 5g-i show. The mean value of position errors is zero by definition of the similarity transformation, and the standard deviations of position errors for this standard array are in Table 1. We revisit these quantities to clarify their meaning.

**Table 1. Standard deviation of position errors from widefield measurements**

| Array | x direction (nm) | y direction (nm) |
|---|---|---|
| Standard process | 1.95 ± 0.03 | 1.97 ± 0.03 |
| Low current, long dwell | 2.43 ± 0.04 | 2.00 ± 0.03 |
| Low current, many passes | 2.11 ± 0.04 | 1.35 ± 0.02 |

Uncertainties are one standard error of the standard deviation.[50]

**Z position**

Optimal use of the aperture array requires control of its z position with respect to the imaging system, and, by extension, its orientation around the x and y axes[51]. Although our nominal depth of field of nearly 1 µm is much greater than our positioning resolution in the z direction of 10 nm, position errors in the x and y directions are still sensitive to changes in the z direction that are as small as 10 nm, which deform the imaging field radially, as Supplementary Figs. S13 and S14 show. For z positions beyond 150 nm from optimal focus, the standard deviation of position errors increases by more than 1 nm. Correction of experimental data will typically require disengagement of a reference material and engagement of an experimental sample, which can cause localization errors from variation in z position. This sensitivity also indicates the importance of microscope stability, as we investigate subsequently.

**Scanning measurements**

To validate our widefield measurements and correction of position errors, we scan the aperture array to sequentially position all apertures that comprise the data in Figure 5 within the central 100 µm$^2$, or 0.2 %, of the imaging field area. This scanning measurement minimizes the effects of photon-optical aberrations to the extent that we can sample them with an array pitch of 5 µm, as Figures 4 and 5d-f show. Pitch values within unit cells of the array are independent of the resolution and repeatability of the scanning stage of the optical microscope. For 1 600 pairs of apertures, scanning measurements result in pitch values that are apparently consistent with widefield measurements, as Supplementary Table S6 shows.

This consistency is only superficial, however, as a deeper analysis shows that scanning and widefield measurements each include multiple sources of error and enables discrimination between the errors. Further details are in Supplementary Note S8. From this analysis, we determine that placement precision results in position errors with a standard deviation of 1.71 nm ± 0.05 nm in the x direction and 1.81 nm ± 0.05 nm in the y direction[52], and that widefield measurements have a localization error of 0.62 nm ± 0.20 nm in the x direction and 0.72 nm ± 0.19 nm in the y direction, independently of empirical localization precision. These uncertainties are standard errors. Further details are in Supplementary Table S7.

Virtually all measurements have errors that limit accuracy at some scale, and our quantification of localization error in widefield measurements is an important advance. One metric for assessing the resulting performance is the field size to localization error ratio of $3 \times 10^5$. To our knowledge, this is the best accuracy for a localization measurement in widefield optical microscopy.

**Chromatic aberrations**

Registration of localization data from different wavelengths can result in errors from chromatic aberrations. To study these effects, we sequentially transilluminate the aperture array with three colors, acquiring three micrographs at each z position. For each color, we determine the z position of the optimal focal plane, the mean value of image pixel size, and the correction model. The mean values of image pixel size differ due to lateral chromatic aberration, and the z positions of the optimal focal planes differ due to axial chromatic aberration, as Supplementary Table S8 shows.

The difference in mean values of image pixel size, and a lateral offset, dominate registration errors, as Figure 7a-c shows for peak wavelengths of 500 nm and 630 nm. We reduce the effects of axial chromatic aberration by selecting and registering micrographs at the optimal focal plane for each color. Registration errors increase for a common z position for multiple colors due to defocus of at least one color, as Supplementary Fig. S15 shows. A similarity transform of the localization data before registration reduces the errors in Figure 7a-c, resulting in systematic errors from the dependence of distortion on color, extending to over 15 nm, as Figure 7d-f shows. Previous studies have empirically modeled such errors without characterizing the contributing effects[22, 23, 25, 26]. These errors are due only to chromatic aberrations, adding to the errors in Figure 5. In a novel analysis, we correct the data from each color prior to the similarity transform. This correction removes the systematic errors from Figure 5a-f and Figure 7d-f, resulting in registration errors that are apparently random, as Figure 7g-i shows. The corresponding localization errors are 0.35 nm ± 0.01 nm in the x direction and 0.47 nm ± 0.01 nm in the y direction. These uncertainties are standard errors. These localization errors are consistent with but smaller than the localization error that we determine from a comparison of widefield and scanning measurements, indicating the existence of systematic components of localization error that cancel in data registration. Further details and the registration of other colors are in Supplementary Fig. S16, Supplementary Note S9, and Supplementary Tables S9 and S10.

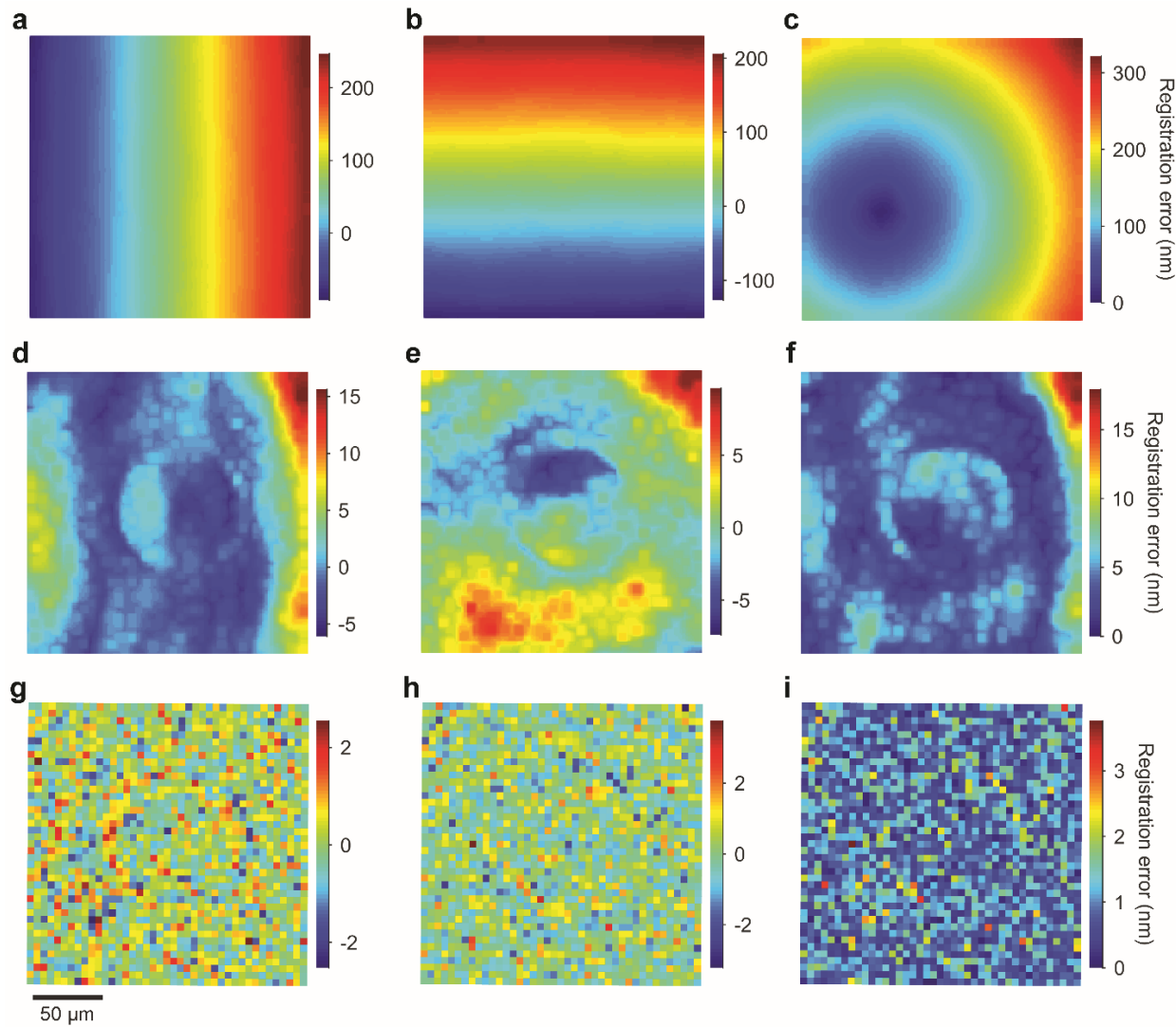

**Figure 7.** Registration errors. **(a-c)** Plots showing registration errors in (a) the x direction, (b) the y direction, and (c) total magnitude, due mostly to different mean values of image pixel size and a lateral offset for localization data of different colors. **(d-f)** Plots showing registration errors in (d) the x direction, (e) the y direction, and (f) total magnitude, after applying a similarity transformation to the localization data, due mostly to variable distortion from chromatic aberration. **(g-i)** Plots showing registration errors in (g) the x direction, (h) the y direction, and (i) total magnitude, after applying correction models to the localization data before a similarity transformation, due mostly to localization error and empirical localization precision.

**Emission source**

We compare transillumination of empty apertures[23] and epiillumination of fluorescent dye in the apertures[27]. The emission wavelengths are similar but not identical for this comparison, as Supplementary Fig. S6 shows. As an exemplary quantity for comparison, the mean values of image pixel size are 100.07 nm for transillumination and 100.16 nm for epiillumination, which differ by more than is attributable to any potential effects of chromatic aberrations, as Supplementary Table S8 shows. These results indicate effects of the illumination and aperture optics, and the requirement for matching the emission of light from apertures to an experimental system to calibrate it. Our reference material and calibration method work equally well for either experimental configuration, indicating their general applicability, as Supplementary Fig. S17 shows. Diverse sample environments are relevant to localization microscopy, motivating future studies of their effects on fluorescence emission and microscope calibration.

**Critical dimensions**

We have assumed the absence of effects of electron-optical aberrations on placement accuracy, which would corrupt calibration of systematic effects of photon-optical aberrations. We test this possibility in two ways. First, because the lateral extent of the aperture array exceeds that of the imaging field, we can independently measure different regions of the array. If electron-optical aberrations were significant, then the photon-optical correction would erroneously include their effects at the array center, resulting in systematic errors upon application of the correction to other regions. No such errors are apparent, as Supplementary Fig. S14 shows. Second, we sample the full extent of the aperture array by scanning 100 pairs of apertures through the central 0.2 % of the imaging field area. No systematic variation in pitch from electron-optical aberrations is apparent, as Supplementary Fig. S18 shows.

In a novel test of placement accuracy, we pattern an independent aperture array using a second lithography system. Widefield measurements reveal that the two arrays differ in mean pitch by 0.01 pixels or approximately 1 nm, as Supplementary Table S11 shows. This difference is extremely statistically significant, with a p-value of 0.0006 for the x direction and 0.0004 for the y direction, but exceeds the position resolution of the lithography stages by less than a factor of two and is approximately half of the standard deviation of position errors due to placement precision. This analysis provides an estimate of placement accuracy, with a corresponding systematic error of image pixel size of 1 nm / 5000 nm = 0.02 %. Importantly, such errors sum arithmetically with distance, as Figure 5a-f shows, so that placement accuracy ultimately limits localization accuracy[28]. However, this limitation of the reference material results in a relative error of only 0.02 % in our analysis of placement precision and empirical localization precision. To our knowledge, this is the most rigorous analysis of a reference material for localization measurements across a wide field.

Our new measurement capability closes the gap between common optical microscopes and uncommon instruments for dimensional metrology,[53] and is immediately applicable to new tests of aperture arrays. For example, using widefield measurements, we can rapidly quantify the dependence of placement precision on fabrication parameters such as dose rate. We decrease the electron-beam current and increase the dwell time by a factor of five with respect to the standard process. The standard deviation of position errors in the x direction increases, as Table 1 and Figure 8a-c show, indicating an asymmetry of our lithography system and that placement precision degrades with decreasing dose rate. Second, we reduce the dwell time by a factor of eight, and overwrite the pattern eight times. The standard deviation of position errors decreases in the y direction, but systematic effects increase this value in the x direction, as Table 1 shows, and a stripe pattern emerges, as Figure 8d-f shows. This pattern further indicates an asymmetry of our lithography system and that aperture placement errors compound with pattern overwriting. Interestingly, regions of Figure 8d,f show systematically smaller position errors, indicating a useful anomaly of the patterning process. These results are all roughly consistent with the specification of beam positioning of 2 nm for our lithography system, but manifest unpredictable irregularities. The high speed and low cost of critical dimension localization microscopy would facilitate quality control of aperture arrays in their production as reference materials.

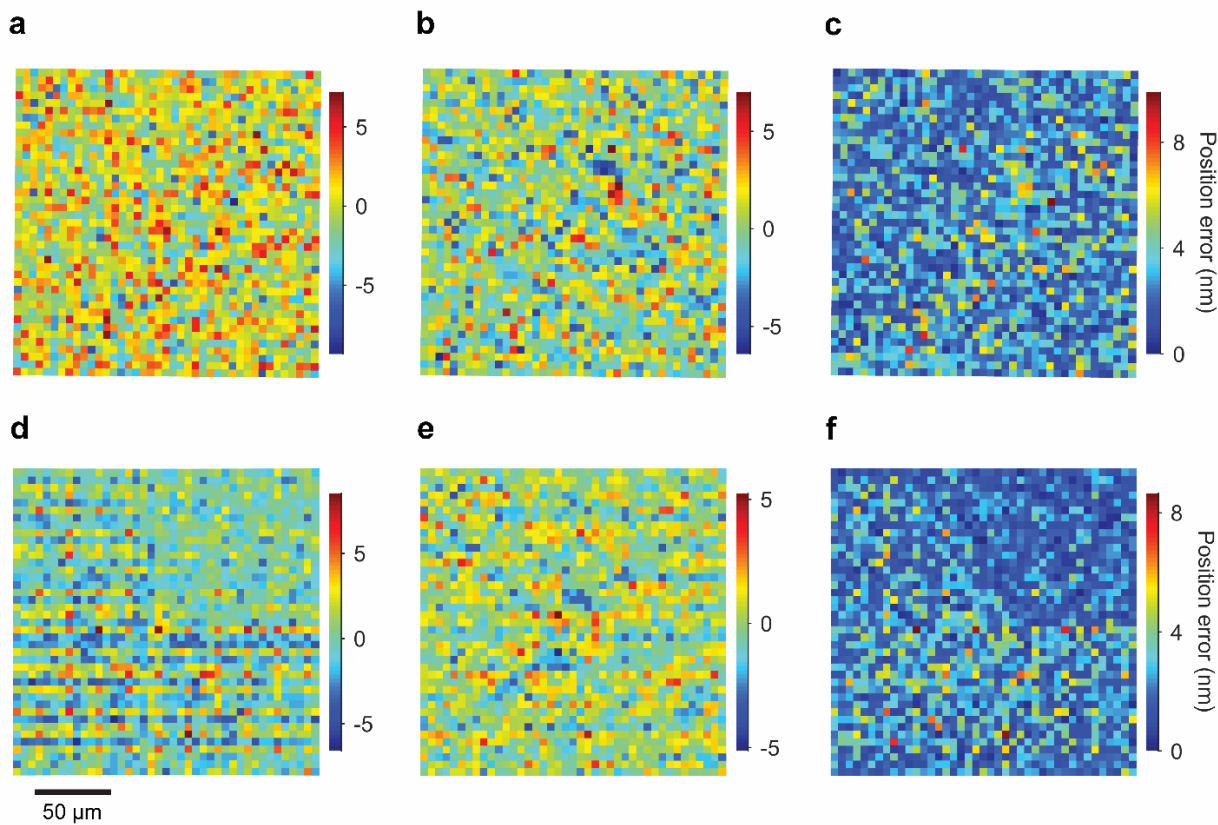

**Figure 8.** Patterning processes. (**a-c**) Plots showing position errors in (a) the x direction, (b) the y direction, and (c) total magnitude, after correcting measurements of aperture positions from an array that we pattern by decreasing the electron-beam current from 1.0 nA to 0.2 nA and increasing the dwell time proportionately to deliver the same dose. (**d-f**) Plots showing position errors in (d) the x direction, (e) the y direction, and (f) total after correcting measurements of aperture positions from an array that we pattern by decreasing the electron-beam current from 1.0 nA to 0.125 nA, maintaining the dwell time, and taking eight passes to deliver the same dose.

**Nanoparticle fiducials**

Transillumination of the aperture array produces an array of point sources that are static with respect to the imaging substrate at any scale that is relevant to our measurements, providing a stable reference material for evaluating any apparent motion of fluorescent nanoparticles as fiducial markers. We localize apertures or nanoparticles in an image series, and assess the apparent motion of each point source using two-dimensional rigid transformations to register corresponding points in image pairs. We quantify apparent motion as the standard deviation of the registration errors over $\sqrt{2}$. Further details are in Supplementary Note S10. This analysis eliminates unintentional motion of the measurement system in the x and y directions, but not in the z direction, as a source of error. For static point sources of one color, registration errors are due only to empirical localization precision and random components of localization error. Normalization of this value by theoretical localization precision allows for direct comparison of nanoparticles and apertures. The aperture array then allows for assessment of additional apparent motion. Any such motion of nanoparticles that exceeds that of apertures is due to actual motion. In this evaluation, the time that is necessary for our microscope to image through focus provides an experimental boundary between faster and slower time scales.

Rigid registration of consecutive images enables tests of motion at a time scale of $10^{-1}$ s. Apertures show apparent motion that ranges from 0.30 nm to 0.65 nm in a single lateral dimension, or a factor of 1.2 to 2.0 times the Cramér–Rao lower bound for each aperture, as Supplementary Fig. S19 shows. For fluorescent nanoparticles on a microscope coverslip, apparent motion ranges from 0.30 nm to 0.85 nm, or a factor of 1.2 to 1.9 times the Cramér–Rao lower bound for each nanoparticle, as Supplementary Fig. S19 shows. These values exceed the Cramér–Rao lower bound by amounts that are consistent with random components of localization error, demonstrating that the nanoparticles do not move in any way that we can measure at this time scale.

Rigid registration of each image with respect to the first image extends the time scale to $10^1$ s. At this time scale, apertures appear to move radially, with registration errors that increase with distance from the center of the field, as Figure 9 shows. Imaging through focus results in apparent motion[54] that is qualitatively similar, as Supplementary Fig. S20 shows, indicating that this apparent motion is consistent with unintentional motion of the measurement system in the z direction.

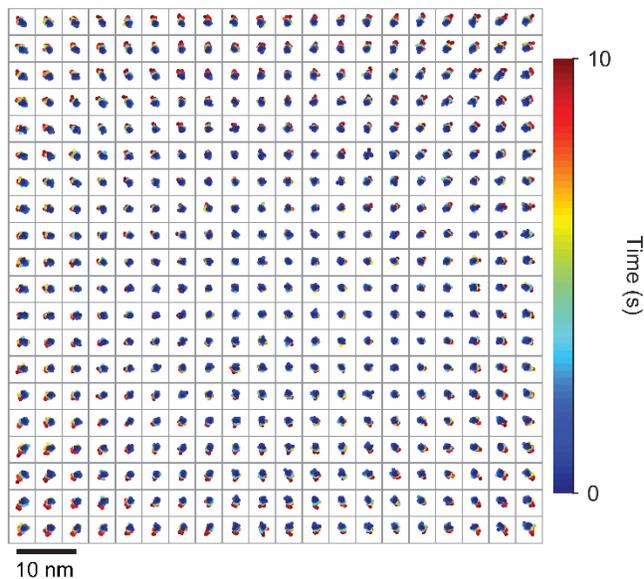

**Figure 9.** Apparent motion. Grid of scatterplots, each corresponding to a single aperture, showing apparent radial motion due to unintentional motion of the measurement system in the z direction over $10^1$ s. The grid spacing indicates an aperture array pitch of 10 µm. The scale bar corresponds to the scatterplots.

At slower time scales, imaging through focus decreases unintentional motion in the z direction to less than 10 nm. Selection of the z position that minimizes registration error, as Supplementary Fig. S8 shows, complements other active[39] and passive[47] methods for mitigating instability of z position. Over $10^4$ s, both apertures and nanoparticles exhibit apparent motion that is quantitatively consistent within their respective mean values of empirical localization precision of approximately 0.43 nm for apertures and 0.55 nm for nanoparticles, as Supplementary Fig. S21 shows. This apparent motion is likely due to differences in z position that are below the positioning resolution between images. Considering that the apertures are static, we conclude that the nanoparticles are static.

These results introduce a new capability for answering open questions about the apparent motion of fluorescent nanoparticles relative to imaging substrates. For an experimental system that is representative of common practice, in that it makes use of typical materials and methods and nonspecific binding, we find that fluorescent nanoparticles can function as fiducial markers with subnanometer stability for several hours. Previous studies reporting nanoparticle motion have not fully characterized the interactions of the components of the measurement system, in particular, unintentional motion along the optical axis, using a stable reference material such as an aperture array. It is evident from our study that this source of motion of any fiducial is clearest across a wide field and upon comparison with other fiducials in an array and is less apparent across a smaller field or at the field center.

**CONCLUSIONS**

It is remarkable that the optical microscope, which has for centuries enabled observations at the micrometer scale, can potentially enable localization measurements at the atomic scale across a millimeter field. In such measurements, localization precision is largely a function of emitter intensity and stability, but localization accuracy depends on a comprehensive calibration of the parts of a measurement system and their interaction. Such calibration is rarely, if ever, implemented, which can cause gross overconfidence in measurement results with small statistical uncertainties but large systematic errors that vary across an imaging field. Such false precision is becoming increasingly problematic as measurements achieve empirical localization precision at the nanometer scale, imaging fields extend into the millimeter scale, and multifocal[55] and multicolor[56] methods emerge to exploit such fields. In this article, we have revealed the surprising extent of this widespread problem and presented a practical solution to it, advancing the practice of localization microscopy.

We have developed the aperture array into a multifunctional reference material that is usefully accurate, precise, planar, and stable. By a combination of widefield and scanning measurements, we have calibrated our microscope system and characterized our aperture arrays. For the first time, we have demonstrated subnanometer localization error across a submillimeter field, for multiple colors and emission sources. This new capability has enabled two novel applications. First, critical dimension localization microscopy facilitates rapid characterization of aperture arrays by widefield imaging, allowing for the study of nanofabrication processes and quality control of reference materials for microscope calibration. Second, we exploit the stability of aperture arrays to evaluate the stability of nanoparticle fiducials, which multiple studies have called into question. We find that microscope instability can obscure the true stability of fluorescent nanoparticles on an imaging substrate, and we provide a method for evaluating different systems.

Our study motivates future work including characterization of aperture arrays by other forms of critical dimension metrology, integration of aperture arrays with various sample environments, and fabrication of other types of reference materials for localization microscopy.

**ACKNOWLEDGMENTS**

The authors acknowledge Glenn Holland for designing and fabricating the stage insert, Kerry Siebein for performing electron microscopy of aperture arrays, Andras Vladar and Stuart Stanton for reviewing and commenting on the manuscript, and four anonymous reviewers for providing constructive comments. The authors acknowledge support of this research under the National

Institute of Standards and Technology (NIST) Innovations in Measurement Science Program, the NIST Center for Nanoscale Science and Technology, and the NIST Physical Measurement Laboratory. CRC acknowledges support under the Cooperative Research Agreement between the University of Maryland and the NIST Center for Nanoscale Science and Technology, award number 70ANB10H193, through the University of Maryland.

**AUTHOR CONTRIBUTIONS**
JG and SMS conceived and obtained funding for the study. CRC and SMS designed the experiments. BRI fabricated aperture arrays. CRC and VAA collected data. CRC and CDM designed analysis algorithms. CRC, JG, and SMS analyzed data. CRC and SMS wrote the manuscript. CDM, JG, VAA, JAL, and BRI reviewed and edited the manuscript.

**AUTHOR NOTES**
The authors declare no conflicts of interest. Further details of the experimental system are available from the corresponding author upon reasonable request.

# Supplementary information *for*
# Subnanometer localization accuracy in widefield optical microscopy

Craig R. Copeland[1, 2], Jon Geist[3], Craig D. McGray[3], Vladimir A. Aksyuk[1], J. Alexander Liddle[1], B. Robert Ilic[1], and Samuel M. Stavis[1, *]

[1]Center for Nanoscale Science and Technology, National Institute of Standards and Technology, Gaithersburg, Maryland 20899, [2]Maryland NanoCenter, University of Maryland, College Park, Maryland 20742, [3]Engineering Physics Division, National Institute of Standards and Technology, Gaithersburg, Maryland 20899, [*]sstavis@nist.gov

**INDEX**

**Note S1.** Aperture array – fabrication

We begin with silica substrates with manufacturer specifications of thickness of approximately 170 µm, surface roughness of less than 0.7 nm root mean square, scratch number of 20, dig number of 10, flatness deviation from $2.5\times10^{-4}$ nm·nm$^{-1}$ to $5.0\times10^{-4}$ nm·nm$^{-1}$, and a parallelism of better than 0.15 mrad. We deposit a titanium film with a thickness of approximately 10 nm as an adhesion layer, a platinum film with a thickness of approximately 80 nm for optical opacity, a positive-tone electron-beam resist film with a thickness of approximately 120 nm, and an aluminum film with a thickness of approximately 15 nm for charge dissipation.

We use two electron-beam lithography systems, enabling comparison of independent aperture arrays to test placement accuracy, and fabrication of different types of aperture arrays that use and test the different operating modes of the systems. Other than different load locks, the lithography systems have nearly identical hardware. Each system has a scanning stage with two laser interferometers to measure stage position in the x and y directions. The resolution of a stage position measurement is 632.8 nm / 1024 = 0.6180 nm, with traceability to the SI through the operating wavelength of the helium–neon laser. One lithography system operates four of five electron-optical lenses and has a write field of 1 mm by 1 mm, which is useful to avoid stitching errors in patterning aperture arrays for widefield imaging, and has a specification for beam placement of 2 nm. The electron-beam current for this system is typically 1.0 nA, although we reduce it in some tests of patterning parameters that we note. The other lithography system operates five of five electron-optical lenses and has a better specification for beam placement of 0.125 nm, which nominally improves placement precision, but does so over a smaller write field of 62.5 µm by 62.5µm. The electron-beam current for this system is 1 nA. We perform a Monte Carlo simulation of electron trajectories in the film stack to correct the pattern data for proximity effects at an accelerating voltage of 100 kV, and we fracture the pattern data into polygons.

After electron-beam exposure, we remove the aluminum film with tetramethylammonium hydroxide and cold-develop the electron-beam resist in hexyl acetate. Finally, we mill the apertures with argon ions, using a secondary-ion mass spectrometer to monitor emission products and stop at the top surface of the silica substrate. The electron-beam resist is not trivial to remove after argon-ion milling and does not affect the function of the aperture array in any way that we are aware of, so we leave the resist in place.

Further characteristics of aperture arrays are in Table S1.

**Table S1.** Aperture array – characteristics

| Array pitch (µm) | Array extent (µm) | Nominal aperture diameter (nm) | Point spread function width[*] (pixels) |
|---|---|---|---|
| 5 | 350 by 350 | 200 | 1.28 ± 0.03 |
| 5 | 350 by 350 | 300 | 1.24 ± 0.02 |
| 5, 10 | 350 by 350 | 400 | 1.27 ± 0.02 |
| 5 | 350 by 350 | 500 | 1.37 ± 0.01 |
| 5 | 62.5 by 62.5 | 500 | 1.39 ± 0.01 |

[*]We characterize the width of the point spread function as $(\sigma_x + \sigma_y)/2$, as Note S7 describes in more detail. Uncertainties are one standard deviation. The mean size of image pixels is approximately 100 nm.

**Note S2.** Aperture array – characterization

We inspect the standard aperture array by scanning electron microscopy, as Fig. S1 shows, at an accelerating voltage of 1 kV and using an Everhart-Thornley detector at a working distance of 9 mm. The apertures are approximately circular with shape irregularity at the scale of tens of nanometers and nonvertical sidewalls, resulting in functional diameters at the silica surface that are apparently smaller than the nominal diameters. We do not attempt to quantify array pitch from these electron micrographs. To do so at the relevant scale would require calibration of the electron microscope and localization analysis that are beyond the scope of this study.

We measure the upper surface topography of the standard aperture array by interferometric optical microscopy, as Fig. S2 shows, at a peak wavelength of 475 nm with a bandwidth of 125 nm. The z position of the piezoelectric stage of this microscope is traceable to the SI through a reference material for step height, and we further calibrate these measurements using a reference flat of silicon carbide. We extract the center of the interference pattern as a function of z position as the location of the reflecting surface. We fit the resulting upper surface topography of the aperture array to a plane to level it and analyze the z-position variation of the upper surface as an indicator of the lower interface between silica and titanium. We expect and observe scratches and digs consistent with the polish of the silica substrate transferring through conformal films. The standard deviation of z position is 1.76 nm, such that the upper surface is effectively flat within the z-position resolution of 10 nm of our localization microscope. Therefore, in subsequent analysis, we ignore any nonflatness of the aperture array. However, in the production of reference materials for localization microscopy in three dimensions, this issue motivates the use of even flatter substrates, or the characterization and analytical correction of any nonplanar surface topography of the aperture array.

After developing our localization measurements and analyses, we apply them to test the extent to which apertures of varying nominal diameters appear as point sources. We summarize these results in Table S1, and describe them in more detail in Note S7. These results indicate that the apertures have functional diameters that are smaller than their nominal diameters, or that our microscope system does not achieve its expected spatial resolution, or a combination of these two effects.

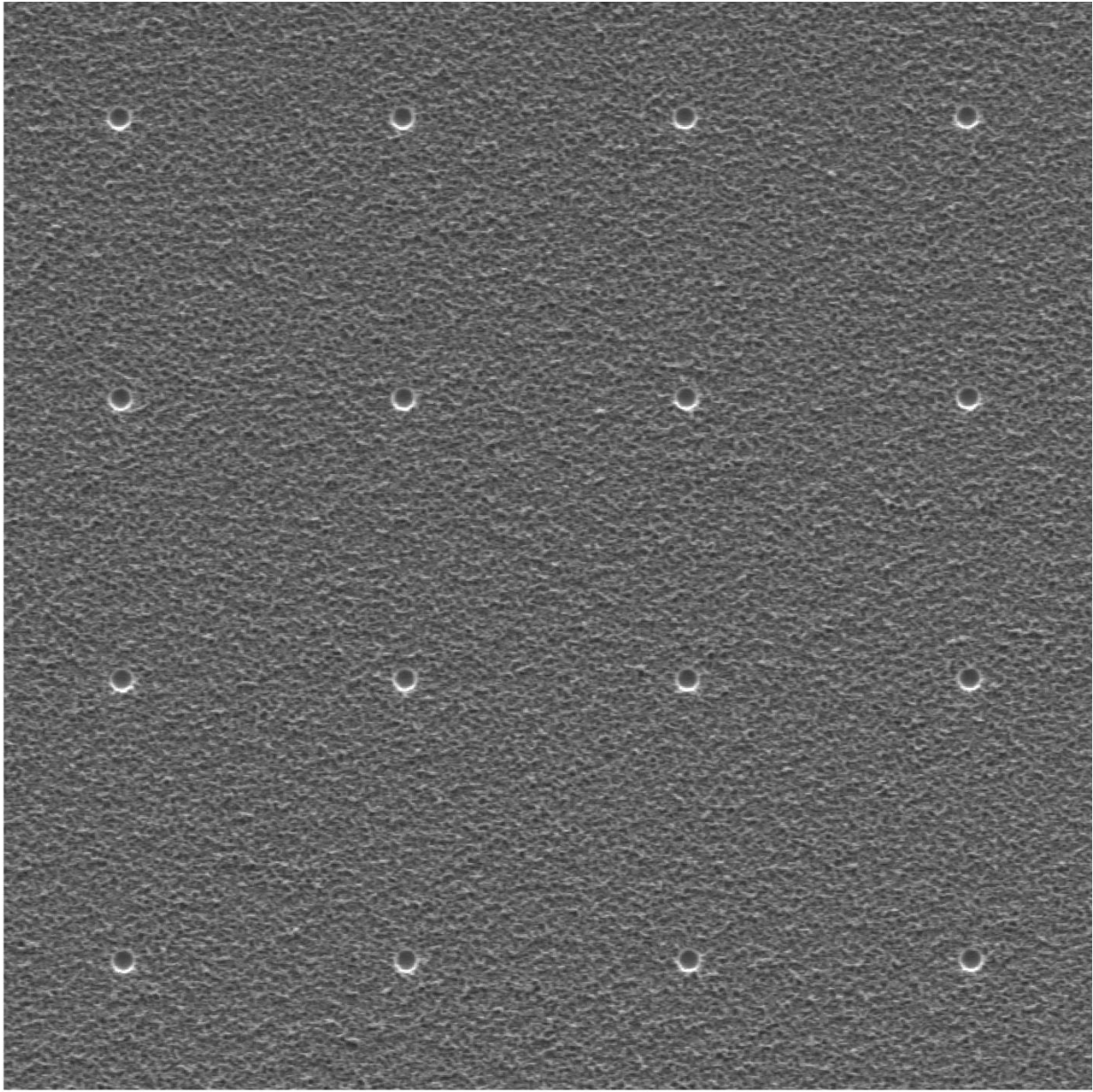

1 μm

**Figure S1.** Aperture array – electron microscopy. Scanning electron micrograph showing 16 apertures. Surface texture around the apertures is from electron-beam resist.

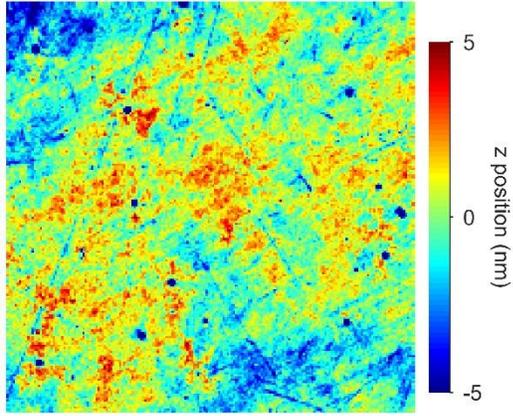

**Figure S2.** Aperture array – interferometric optical microscopy. Interferometric optical micrograph showing the upper surface topography of a representative region corresponding approximately to the aperture array. The apertures are below the resolution of this imaging system. Scratches and digs in the upper surface are consistent with the polish of the lower silica surface. The standard deviation of z position is 1.76 nm.

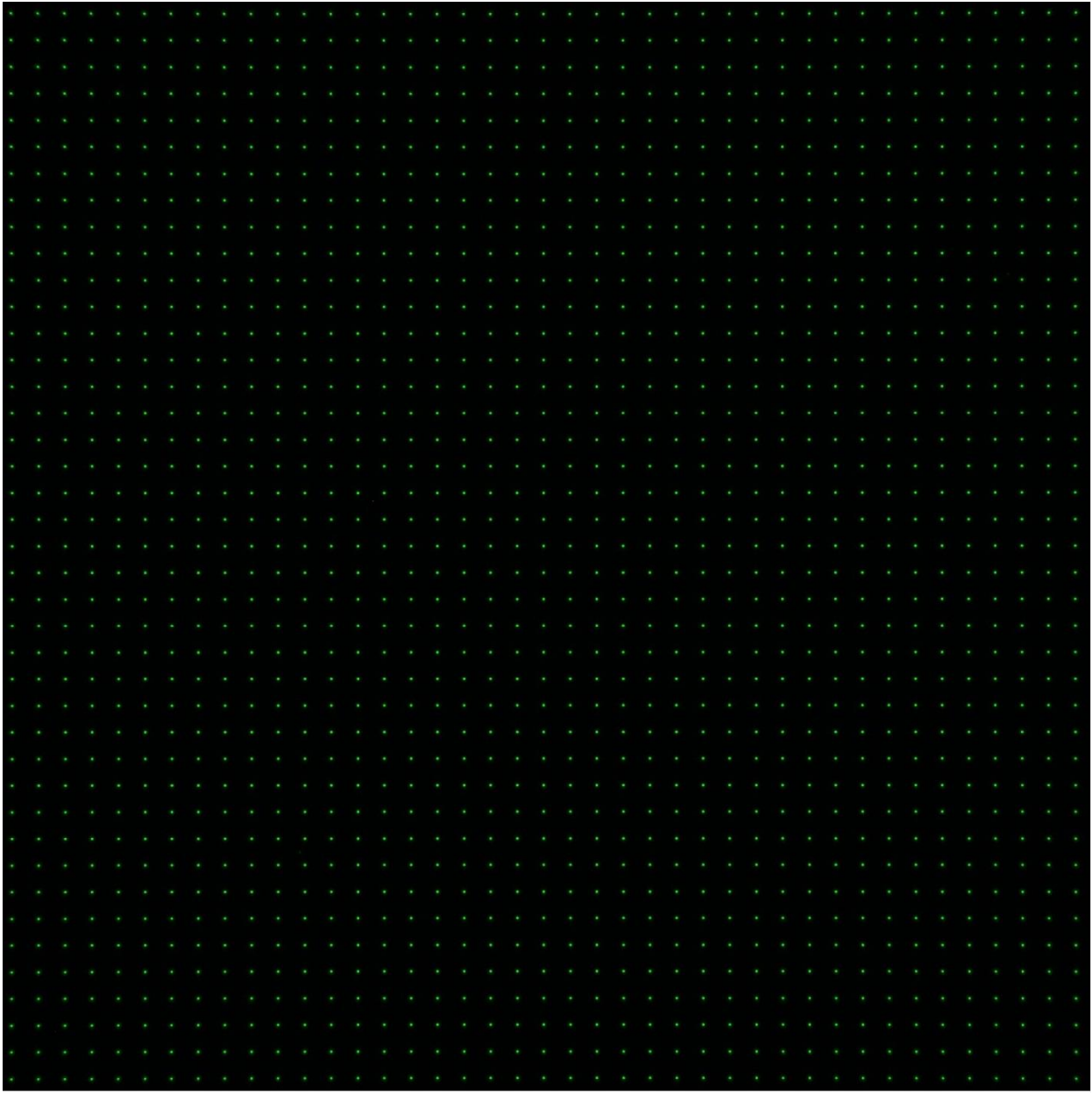

**Figure S3.** Aperture array – optical microscopy. Brightfield optical micrograph showing the transmission of light through an aperture array over the full field of the imaging system of approximately 200 μm by 200 μm. False color represents the illumination wavelengths of around 500 nm.

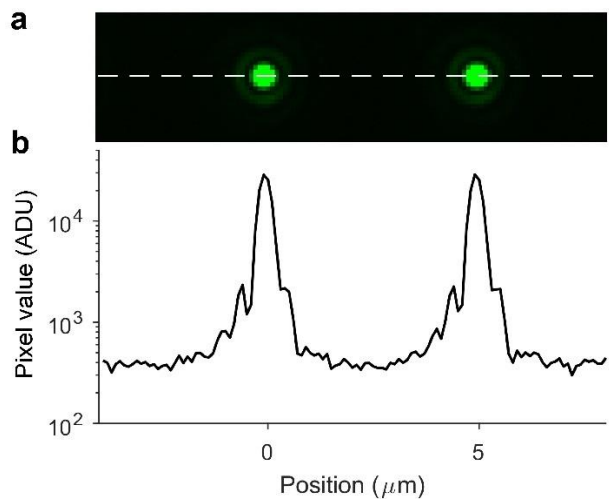

**Figure S4.** Aperture array – point spread functions. **(a)** Brightfield optical micrograph showing the point spread functions from two apertures with nominal diameters of 400 nm in an array with a nominal pitch of 5 µm. **(b)** Plot showing pixel value along the white dashed line in (a). Airy rings are evident on a logarithmic scale for the vertical axis. The point spread function from the left aperture decays to background by approximately 3 µm from the center position of the aperture. This shows that an array pitch of 5 µm provides sufficient separation of adjacent apertures such that their signals do not appreciably overlap within the region of interest for localization analysis, which is approximately 1 µm around the center position of each aperture.

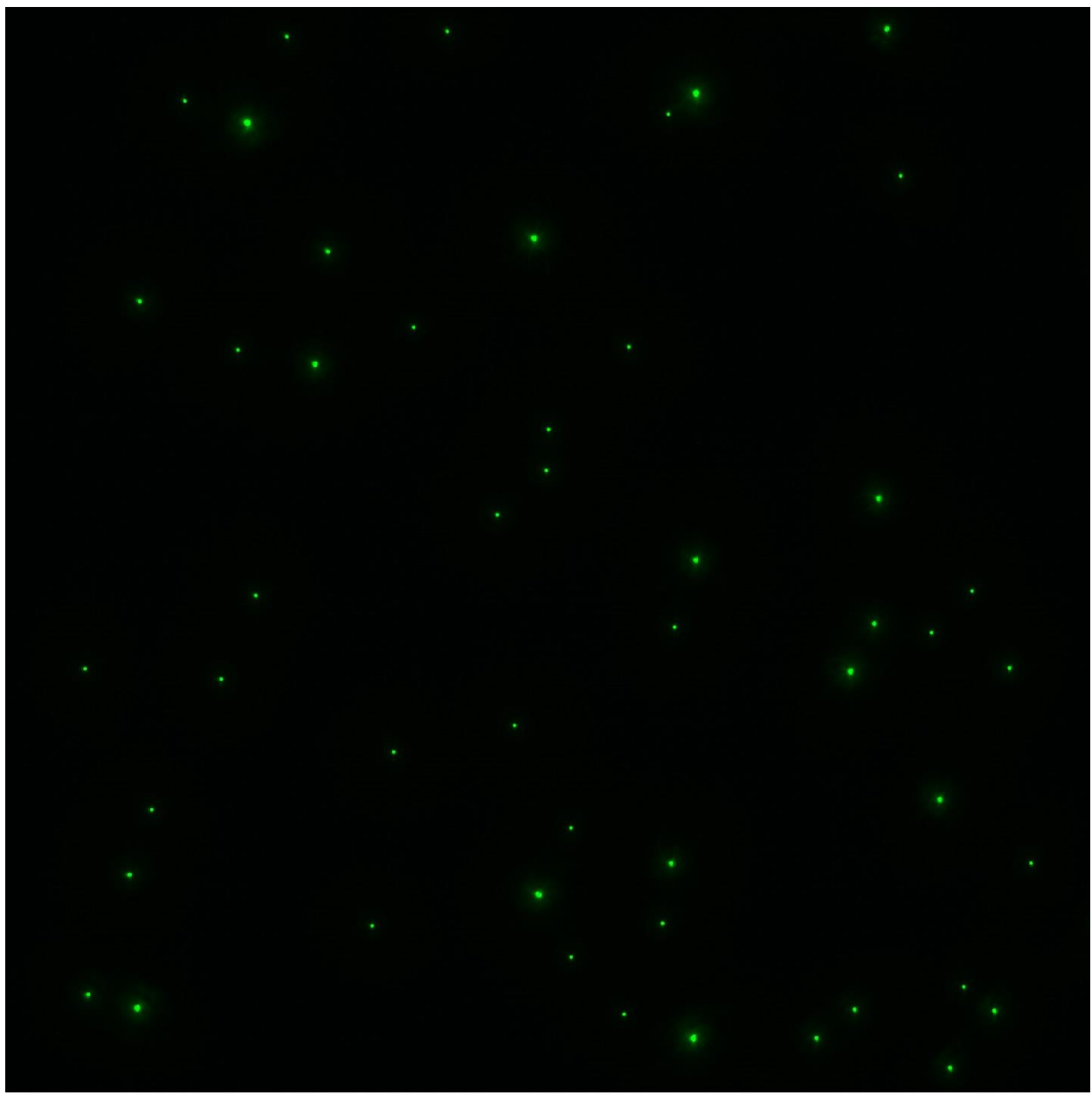

10 µm

**Figure S5.** Nanoparticle fiducials. Fluorescence micrograph showing fluorescent nanoparticles with a carboxylate coating on a borosilicate coverslip with a poly-D-lysine coating. In subsequent analysis, we ignore aggregates of nanoparticles, which are evident as images that are brighter and larger than single point spread functions.

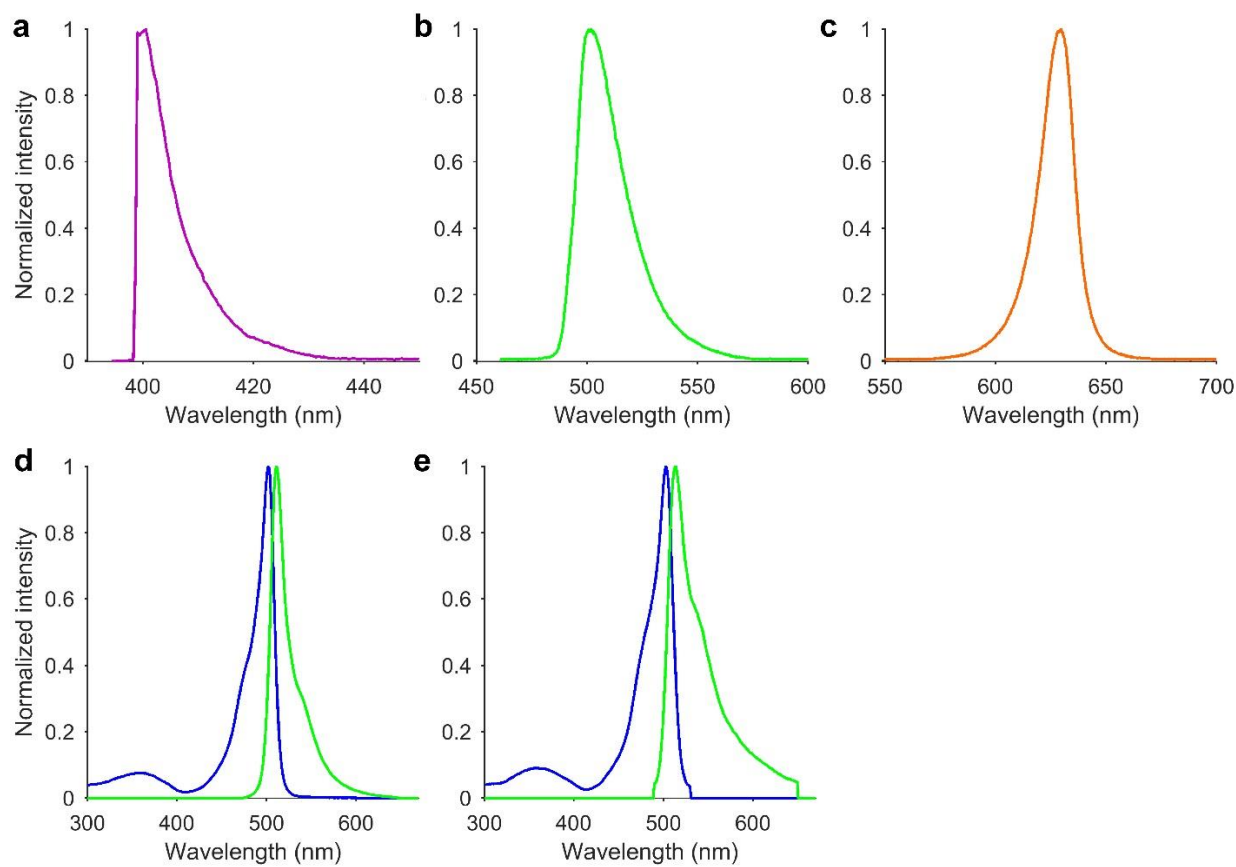

**Figure S6.** LED and dye spectra. (**a**-**c**) Plots showing experimental emission spectra of LED arrays with peak wavelengths of (a) 400 nm, (b) 500 nm, and (c) 630 nm. (**d**-**e**) Plots showing nominal excitation (blue) and experimental emission (green) spectra of (d) boron-dipyrromethene dye in N,N-dimethylformamide solution and (e) in amorphous polystyrene nanoparticles.

**Note S3.** Sample leveling

We can level a sample by aligning its surface normal to the optical axis using two methods. The first exploits piezoelectric actuation and characterization of the z position of the objective lens, as we describe in the main text. The second takes advantage of Zernike theory. Both require a stage insert that enables rotation of the sample about the x and y axes, as Fig. S7a-b shows. In the second method, we analyze spatial maps of $\rho$, as we define in Eq. 1 and Eq. S1, across the field. We fit the maps to a linear combination of Zernike polynomials[1] in real time, finding the optimal orientation which minimizes the coefficients for the first-order Zernike polynomials $Z_1^1$ and $Z_1^{-1}$, which model orientation of the sample about the x and y axes, as Fig. S7b-f show.

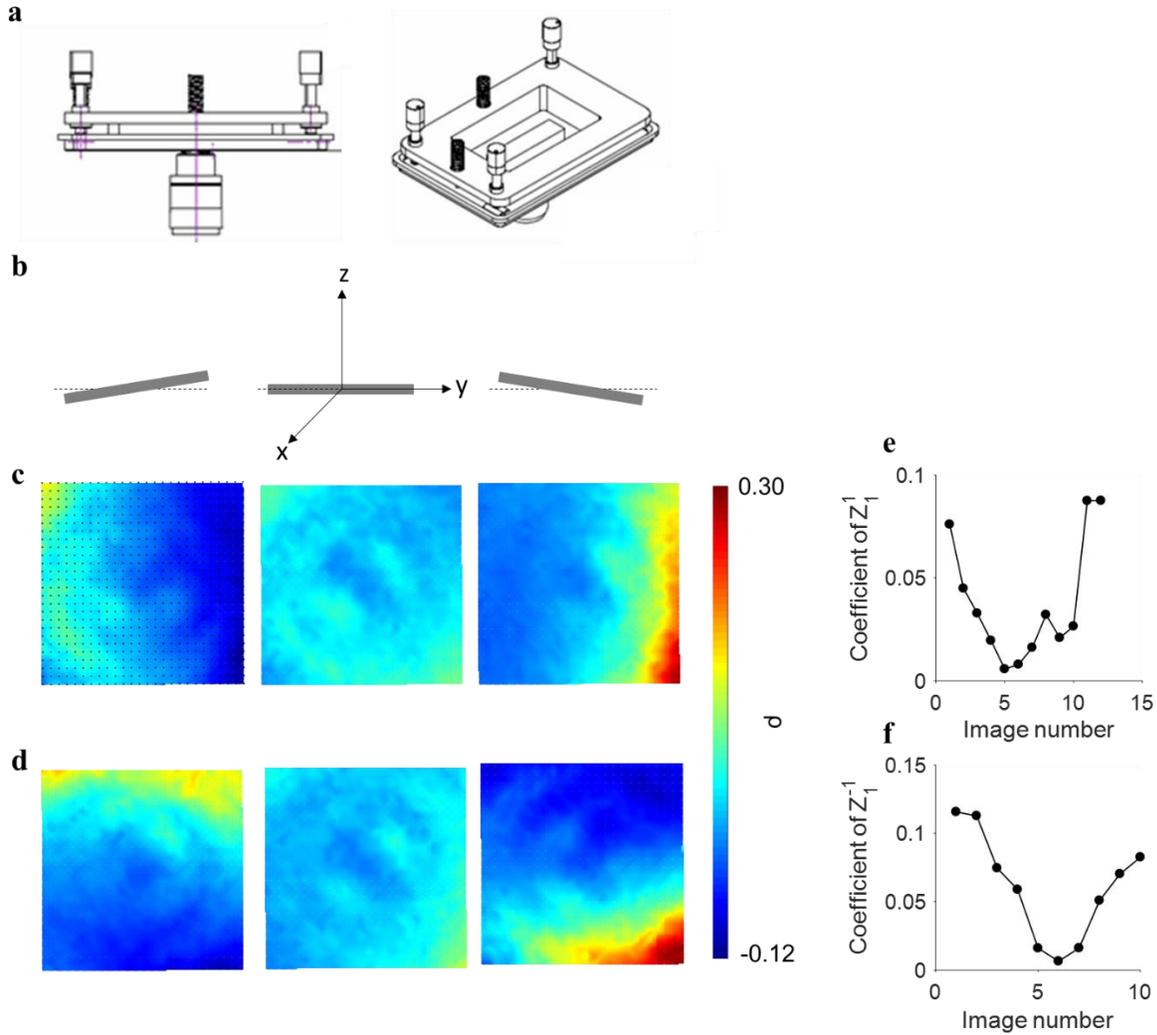

**Figure S7.** Sample leveling. (**a**) Schematic showing sample holder. (**b**) Schematic showing sample orientation about the x axis, not to scale. (**c**) Plots showing $\rho$ at varying magnitudes of orientation about the x axis. Black dots indicate aperture positions. (**d**) Plots showing $\rho$ at varying magnitudes of orientation about the y axis. Orientation direction corresponds to the schematics in (b). (**e**) Plot showing representative values of the coefficient of the Zernike polynomial $Z_1^1$, modeling orientation about the x axis. The minimum corresponds to the center plot in (c). (**f**) Plot showing representative values of the coefficient of $Z_1^{-1}$, modeling orientation about the y axis. The minimum corresponds to the center plot in (d).

**Note S4.** Optimal focus

For any region of interest, from a square micrometer around a single aperture to the full field of the imaging system, we determine optimal focus first by imaging through focus. We then extract the mean amplitude of the point spread functions that are within the region of interest and empirically model the variation of the mean amplitude with respect to z position using a quintic function. We take the maximum value of the model fit as the z position of optimal focus. Fig. S8 shows amplitude as a function of z position for one aperture and mean amplitude as a function of z position for many apertures in one image.

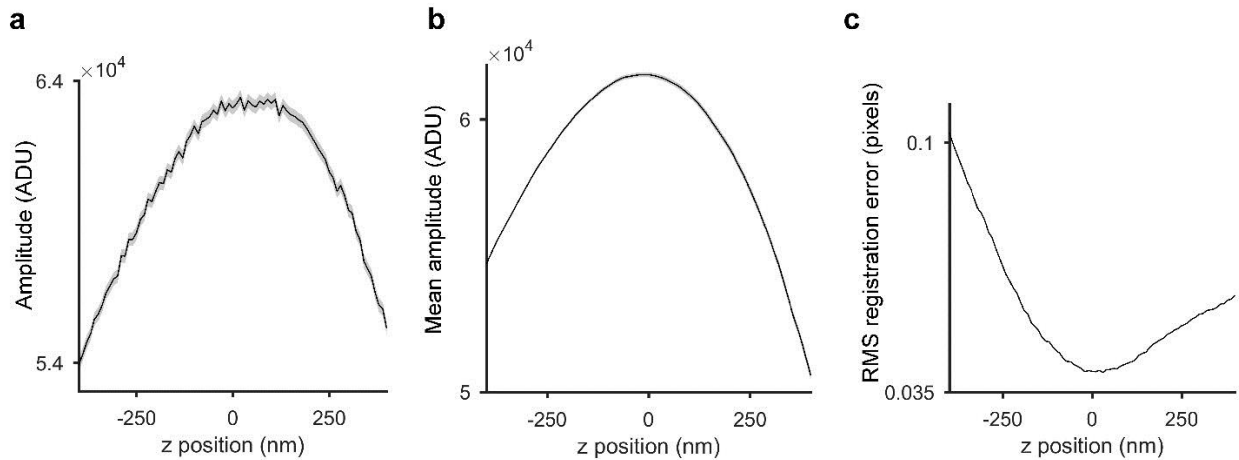

**Figure S8.** Optimal focus. **(a)** Plot showing the amplitude of the point spread function of a single aperture as a function of z position, with a maximum at optimal focus. The grey boundary is one standard deviation. **(b)** Plot showing the mean amplitude of 1600 point spread functions from as many apertures as a function of z position, with a maximum at the optimal focal plane. The z position of optimal focus of the aperture in (a) differs from the z position of the optimal focal plane in (b) due to field curvature. **(c)** Plot showing the root-mean-square error of a rigid registration between images of an aperture array as a function of z position, with a minimum at the z position of the common optimal focal plane between the two images. The grey boundaries in (b) and (c) are one standard error and are comparable in width to the black lines.

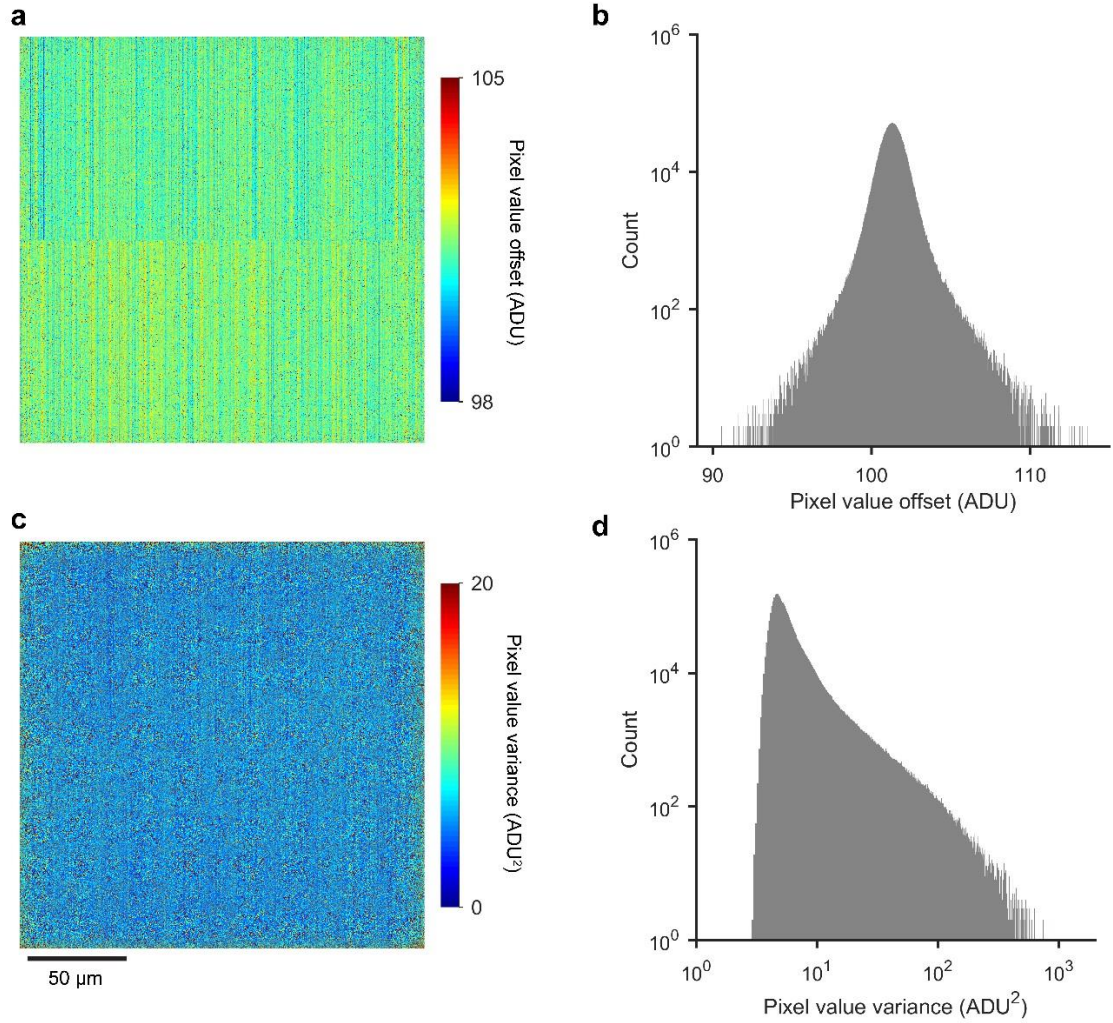

**Figure S9.** Dark calibration of camera. (**a**) Plot showing pixel value offset. (**b**) Histogram showing pixel value offset. (**c**) Plot showing pixel value variance. (**d**) Histogram showing pixel value variance. To clearly show systematic effects in (a) and (c) from the CMOS architecture of the imaging sensor, we restrict the ranges of (a) with respect to (b) and (c) with respect to (d).

**Table S2.** Terminology

| Process | Term | Sources of Error | Quantity |
|---|---|---|---|
| Aperture fabrication | Placement accuracy | Electron-optical aberrations<br>Position resolution of lithography system | Mean magnitude of differences of aperture placements from nominal positions* |
| | Placement precision | Pattern resolution and transfer | Standard deviation of difference of aperture placements from nominal positions* |
| Emitter localization | Theoretical localization precision | Photon shot noise<br>Background noise<br>Image pixel size<br>Point spread function | Cramér–Rao lower bound |
| | Empirical localization precision | Theoretical localization precision<br>Fitting error<br>Unintentional random motion of measurement system | Standard deviation of difference of position measurements from mean value of position measurements |
| Microscope calibration | Position accuracy | Placement precision<br>Photon-optical aberrations<br>Image pixel size<br>Fitting error<br>Unintentional systematic motion of measurement system<br>Empirical localization precision | Position error – difference of aperture position measurement from nominal position* |
| | Correction accuracy | Placement precision | Correction error – difference of placement precision and the standard deviation of position errors in a synthetic array with ideal placement accuracy |
| Error correction | Localization accuracy | Unintentional axial motion of measurement system<br>Correction accuracy<br>Unknown sources of error | Localization error – standard deviation of position errors, independent of placement precision and empirical localization precision |
| Data registration | Registration accuracy | All sources above<br>Chromatic aberration | Registration error – difference of corresponding position measurements from two images |

*Nominal positions are at the nodes of an ideal square array as per our design. Mean differences that do not alter the mean value of array pitch do not affect microscope calibration.

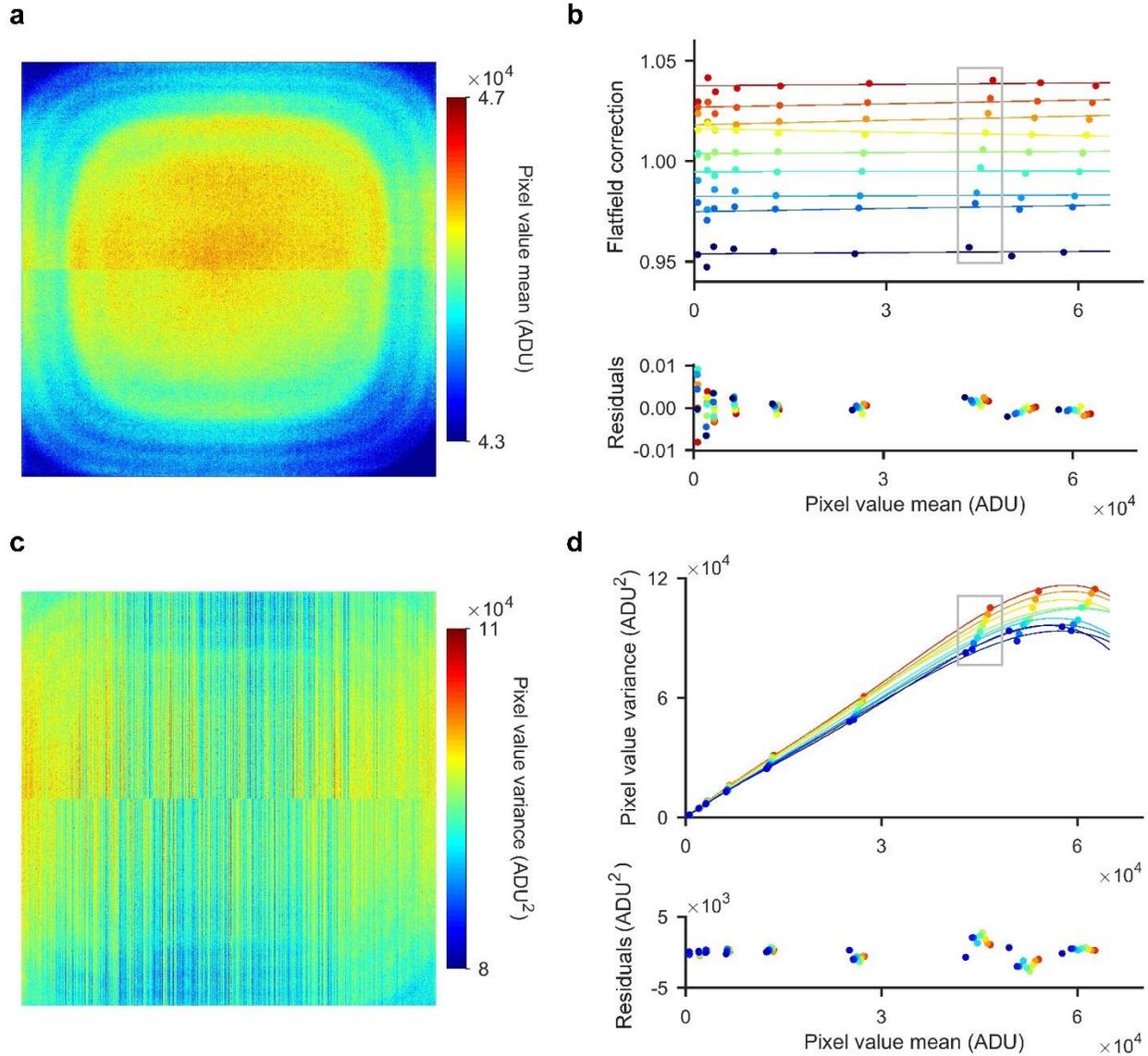

**Figure S10.** Light calibration of camera. (**a**) Plot showing pixel value mean from 15 000 images at one of nine illumination levels. Nonuniformity results from the illumination profile, sensor packaging, and CMOS architecture. (**b**) Plot showing flatfield corrections for nine representative pixels as a function of pixel value mean. The gray box encloses data from the illumination level in (a). The flatfield corrections abruptly increase at low values and then remain nearly constant for the remaining 95 % of the dynamic range. A linear function empirically approximates the flatfield corrections over the full dynamic range. (**c**) Plot showing pixel value variance corresponding to the pixel value mean in (a). Nonuniformity results from sensor packaging and amplifier columns. (**d**) Plot showing pixel value variance, including contributions from shot noise, read noise, and fixed-pattern noise, as a function of pixel value mean for nine representative pixels. The gray box encloses data from the illumination level in (a, c). A quartic polynomial empirically approximates the pixel value variance over the full dynamic range. The ratio of pixel value variance to pixel value mean gives an approximate value of gain. Therefore, the quartic polynomial can provide an estimate of gain for any pixel and pixel value for converting units from ADU to photons, for example, for calculation of a Cramér–Rao lower bound.

**Note S5.** CMOS localization

We test localization accuracy for single emitters over the full dynamic range and field of our CMOS camera. We model the response of each pixel as a Gaussian probability density function, which replaces the Poisson distribution that commonly models shot noise,[2, 3, 4] due to the nonlinear relationship between pixel value and total variance. The probability density function for each pixel incorporates the pixel value offset and flatfield correction in the calculation of the mean or expected pixel value to account for variation in pixel gain, illumination nonuniformity, and the effects of sensor packaging. The variance of the probability density function comes from the quartic function in the main text. We perform Monte Carlo simulations to generate images of a univariate Gaussian point spread function in which this same Gaussian probability density function, incorporating parameter values that correspond exactly to a region of our CMOS camera, determines each pixel value. This analysis results in accurate localization with uncertainties near the Cramér–Rao lower bound, as Table S3 shows for the x direction. We find that using an approximate model for total variance, which includes only contributions from shot noise and read noise for each pixel, results in empirical localization precision and localization accuracy that are equivalent to using the empirical model for the total variance. This demonstrates that, despite the difference between the empirical and approximate variance, which is significant for pixels with values in the top 25 % of the dynamic range, the approximate model is more efficient and is equally accurate even for images of point sources with pixel values that span the full dynamic range of the CMOS sensor.

**Table S3.** CMOS localization

| Number of signal photons | Theoretical localization precision (pixels) | Empirical localization precision* (pixels) | Standard error* (pixels) | Empirical error* (pixels) |
|---|---|---|---|---|
| $4.5 \times 10^5$ | $2.7 \times 10^{-3}$ | $2.9 \times 10^{-3}$ | $4.1 \times 10^{-5}$ | $5.8 \times 10^{-5}$ |
| $7.0 \times 10^5$ | $2.2 \times 10^{-3}$ | $2.4 \times 10^{-3}$ | $3.4 \times 10^{-5}$ | $5.3 \times 10^{-5}$ |

*Values from measurements of 5000 images.

**Note S6.** Localization algorithms

We approximate the point spread function, which varies across the imaging field, with a bivariate Gaussian function,

$$G_{biv}(x, y, \Theta = [A, \sigma_x, \sigma_y, \rho, x_0, y_0, C]) =$$
$$A \cdot \exp - \left( \frac{1}{2(1-\rho^2)} \left[ \frac{(x-x_0)^2}{\sigma_x^2} - 2\rho \frac{(x-x_0)(y-y_0)}{\sigma_x \sigma_y} + \frac{(y-y_0)^2}{\sigma_y^2} \right] \right) + C, \quad \text{(Eq. S1)}$$

where $A$ is the amplitude, $x_0$ is the position of the peak in the x direction, $y_0$ is the position of the peak in the y direction, $\sigma_x$ is the standard deviation in the x direction, $\sigma_y$ is the standard deviation in the y direction, $\rho$ is the correlation coefficient between the x and y directions, and $C$ is a constant background. This model determines the expected pixel value in analog-to-digital units (ADU) for each pixel in an image,

$$E_i(x_i, y_i, \Theta) = G_{biv}(x_i, y_i, \Theta), \quad \text{(Eq. S2)}$$

where $i$ indexes each pixel, $x_i$ is the position of the pixel in the x direction, $y_i$ is the position of the pixel in the y direction. For weighted least-squares, the objective function for fitting this model of the expected pixel values using is,

$$\widehat{\Theta} = \text{argmin} \left[ \sum_i \frac{(I_i - E_i)^2}{gI_i + \sigma_{\text{read},i}^2} \right], \quad \text{(Eq. S3)}$$

where $\widehat{\Theta}$ is the estimate for the parameter set $\widehat{\Theta} = \{A, \sigma_x, \sigma_y, \rho, x_0, y_0, C\}$,

$g$ is the nominal gain of the camera specified by the manufacturer, $\sigma^2_{read,i}$ is the pixel read noise, and $I_i$ is the experimental pixel value after correction for CMOS characteristics,

$$I_i = \frac{I_i^* - o_i}{FF_i}, \quad \text{(Eq.S4)}$$

where $I_i^*$ is the raw pixel value, $o_i$ is the pixel value offset, and $FF_i$ is the flatfield correction. In the case of a Gaussian probability density function for the response of single pixels, the objective function for maximum-likelihood is similar,

$$\hat{\Theta} = \text{argmin}\left[\sum_i \frac{(I_i - E_i)^2}{gE_i + \sigma^2_{read,i}}\right], \quad \text{(Eq. S5)}$$

with the only difference being the replacement of the experimental pixel value $I_i$ in the denominator of Eq. S3 with the model or expected pixel value $E_i$.

If the model systematically underestimates the experimental pixel values, then the presence of the expected pixel value $E_i$ in the denominator of Eq. S5 means that maximum-likelihood gives additional weight to the underestimated pixel, as Fig. 3 shows. In contrast, the presence of $I_i$ in the denominator of Eq. S3 means that weighted least-squares does not have this bias. These effects are the opposite for the case that the model systematically overestimates the experimental values.

We modify our localization algorithm to mitigate such effects. A general solution to this problem of selecting either weighted least-squares or maximum-likelihood is a hybrid objective function, which empirically reduces the effect of model discrepancies whether the model systematically overestimates or underestimates the data,

$$\hat{\Theta} = \text{argmin}\left[\sum_i \frac{(I_i - E_i)^2}{g \cdot \max(I_i, E_i) + \sigma^2_{read,i}}\right], \quad \text{(Eq. S6)}$$

where $\max(I_i, E_i)$ reduces the weight of pixels with significant residuals. Therefore, we term this the light-weighting objective function.

We use unweighted least-squares to determine the starting point for localization with the other algorithms. The field dependence of position estimation with light-weighting, maximum-likelihood, and weighted and unweighted least-squares is in Fig. S11, and a quantitative comparison of empirical localization precision is in Table S4. We derive empirical localization precision from the standard deviation of 100 measurements in an image series of the pitch of each unit cell of the aperture array. The values in Table S4, which average over the x and y directions, are the root-mean-square of the pitch standard deviations over a factor of $\sqrt{2}$ from 1 640 pitches.

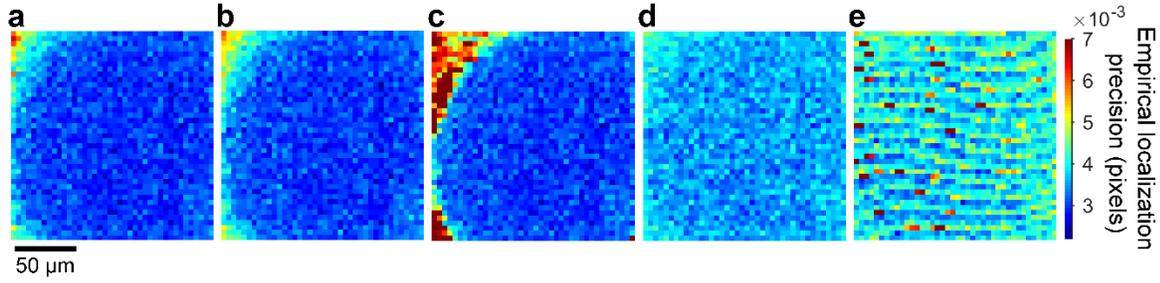

50 μm

**Figure S11.** Localization algorithm performance. (**a-e**) Representative plots showing empirical localization precision across the field for position estimation with (a) light-weighting, (b) weighted least-squares, (c) maximum-likelihood, (d) unweighted least-squares, and (e) light-weighting with a smaller region of interest of 500 nm by 500 nm that excludes much of the point spread function outside of the central peak. The data in (e) is nearly identical for the first three localization algorithms. The mean number of signal photons per point spread function is $5.3 \times 10^5$. For this data, weighted least-squares performs similarly to light-weighting, due to deformation of the point spread function most often causing the model to underestimate the data, but this may not always be the case. Unweighted least-squares generally results in larger uncertainties than the other algorithms and is not suitable for inclusion of CMOS characteristics and shot noise. However, it is also less sensitive to the model discrepancy that Fig. 3 shows, because uniform weighting optimizes the fit to the central peak of the point spread function that is approximately Gaussian. Therefore, unweighted least-squares performs best in field regions with the largest deformation of the point spread function. Similarly, a region of interest that excludes much of the point spread function outside of the central peak results in nearly identical performance of the first three algorithms, but the empirical localization precision is significantly worse overall. The field dependence in (e) indicates systematic effects of pixelation on the definition of a localization region of interest that excludes much of the point spread function outside of the central peak. These results highlight the utility of light-weighting for accommodating deformation of the point spread function. Summary results for the different localization algorithms for different signal intensities and regions of interest are in Table S4.

**Table S4.** Localization algorithm performance

|  | Mean number of signal photons per point spread function | | | |
|---|---|---|---|---|
|  | $5.3 \times 10^5$ | $3.0 \times 10^5$ | $5.3 \times 10^4$ | $5.9 \times 10^3$ |
|  | **Empirical localization precision (pixels)** | | | |
| Light-weighting (Eq. S6) | 0.00295 (0.00398) [*] | 0.00399 | 0.00889 | 0.02710 |
| Weighted least-squares (Eq. S3) | 0.00301 (0.00399) [*] | 0.00391 | 0.00892 | 0.02910 |
| Maximum-likelihood (Eq. S5) | 0.00356 (0.00399) [*] | 0.00795 | 0.01398 | 0.03183 |
| Unweighted least-squares | 0.00339 | 0.00446 | 0.01042 | 0.03165 |

[*]Values in parentheses correspond to a region of interest that includes only the central peak of the point spread function.

**Note S7.** Point source test

We test the extent to which empty apertures with nominal diameters ranging from 200 nm to 500 nm appear as point sources under transillumination. For each value of nominal diameter, we image 400 apertures around the center of the write field and the center of the imaging field. We determine the position of optimal focus as Fig. S8 shows, localize each aperture, extract the standard deviations of the bivariate Gaussian approximation of the point spread function, and evaluate the mean value of $(\sigma_x + \sigma_y)/2$. These values are in Table S1. Apertures with nominal diameters of 200 nm, 300 nm, and 400 nm have equivalent mean values of this quantity, indicating that the functional diameters of these apertures are below the resolution of the imaging system and that they appear as point sources. These mean values of $(\sigma_x + \sigma_y)/2$ exceed the theoretical value of approximately $0.21\lambda/\text{NA} = 90$ nm, likely due to the inclusion of the first Airy ring in the fitting region of interest. Apertures with nominal diameters of 500 nm appear to be slightly larger, indicating that their functional diameters approach the resolution limit of the imaging system. On the basis of this data, in the calibration of our microscope, we typically use apertures with nominal diameters of 400 nm to maximize the number of signal photons.

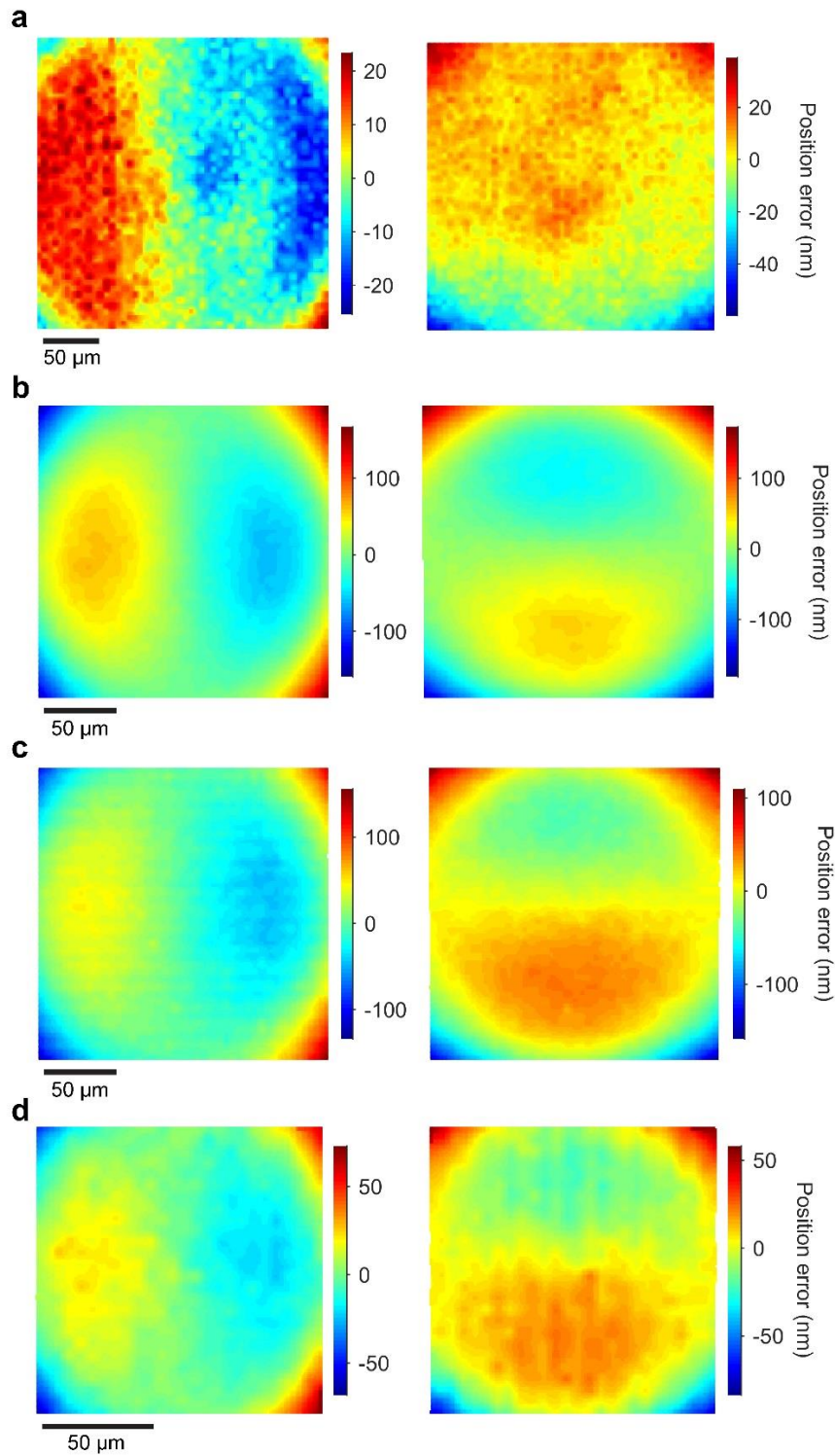

**Figure S12.** Objective lenses. (**a-d**) Plots showing position errors due mostly to using the mean values of image pixel size for four objective lenses with magnification and numerical aperture values of (a) 50× and 0.55, (b) 63× and 1.20, (c) 63× and 1.40, and (d) 100× and 1.46. The left column shows position errors in the x direction. The right column shows position errors in the y direction. We reconfigure the same microscope system for testing each objective lens using an aperture array with nominal diameters of 200 nm or 400 nm. Further specifications of the objective lenses and the resulting standard deviation of position errors are in Table S5. Removing and replacing an objective lens requires recalibration of the microscope. For example, when we remove and replace the objective lens in (b), the mean value of image pixel size changes by up to 0.07 %.

**Table S5.** Objective lenses

| Magnification (×) | Numerical aperture ( ) | Refractive index of immersion medium ( ) | Working distance (mm) | Corrections | Standard deviation of position errors (nm) | |
|---|---|---|---|---|---|---|
| | | | | | x | y |
| 50 | 0.55 | 1.00 | 9.1 | Chromatic, flatfield | 10.85 ± 0.15 | 11.57 ± 0.16 |
| 63 | 1.2 | 1.33 | 0.28 | Coverslip, chromatic, flatfield | 39.95 ± 0.69 | 39.52 ± 0.68 |
| 63 | 1.4 | 1.52 | 0.19 | Coverslip, chromatic, flatfield | 30.53 ± 0.52 | 30.75 ± 0.53 |
| 100 | 1.46 | 1.52 | 0.11 | Coverslip, chromatic, flatfield | 15.64 ± 0.43 | 16.34 ± 0.44 |

All objective lenses are from the same manufacturer. All specifications are nominal values from the manufacturer.

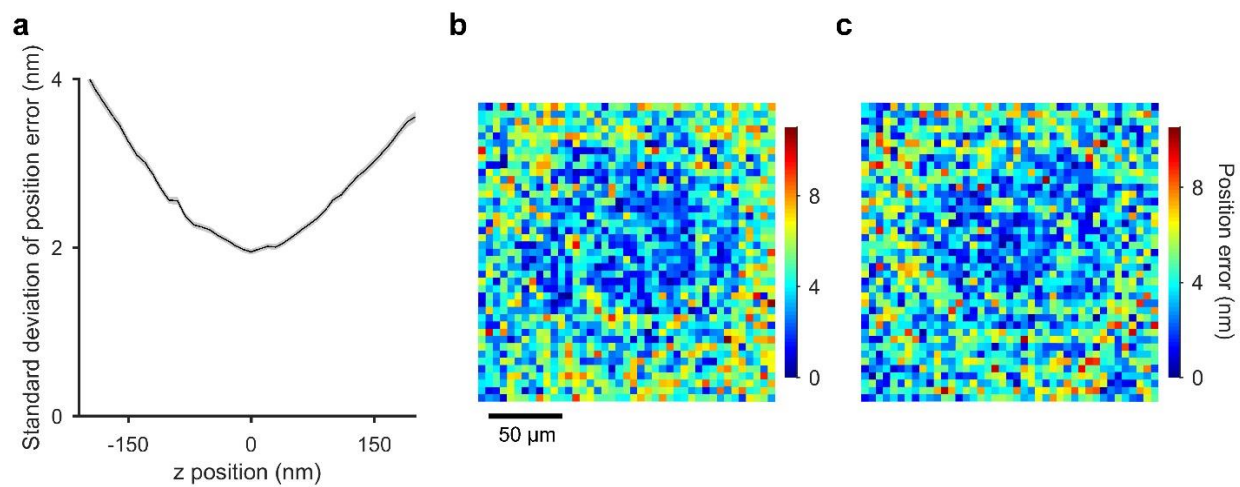

**Figure S13.** Error correction depends on z position. **(a)** Plot showing the pooled standard deviation of position errors in the x and y directions following error correction with respect to z position. The gray boundary is one standard error and is comparable in width to the black line. **(b, c)** Plots showing the total magnitude of position errors at (b) 150 nm below the z position of optimal focus and (c) 150 nm above the z position of optimal focus. Position errors increase with the magnitude of z position away from optimal focus, with a radial deformation of the field.

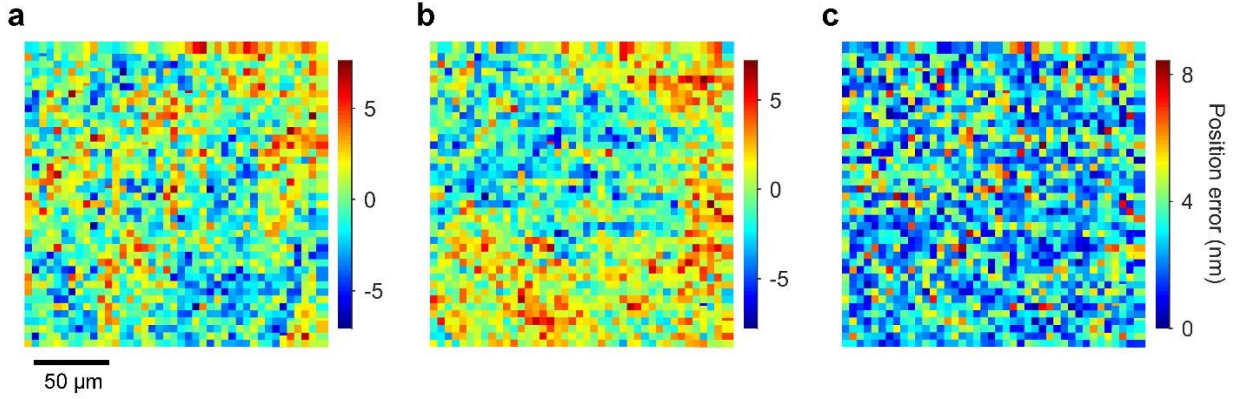

**Figure S14.** Error correction across the aperture array. Plots showing position errors in **(a)** the x direction, **(b)** the y direction, and **(c)** total magnitude, from applying error correction models that we derive from the center of the standard array to a different region of the standard array. Systematic effects in (b) are consistent with variation in z position with respect to the data in Fig. 5 in the main text. Additional systematic effects that may indicate the presence of electron-optical aberrations in the process of electron-beam patterning are not apparent.

**Note S8.** Scanning and widefield measurements

The spatial variances of pitch values across the aperture array from scanning and widefield measurements are, respectively,

$\sigma^2_{\text{pitch},S} = \sigma^2_{\text{lp},S} + \sigma^2_{\text{pp}} + \sigma^2_{\text{le},S}$ (Eq. S7)

$\sigma^2_{\text{pitch},W} = \sigma^2_{\text{lp},W} + \sigma^2_{\text{pp}} + \sigma^2_{\text{le},W}$ (Eq. S8)

where $\sigma^2_{\text{lp},S}$ is the variance from empirical localization precision in scanning measurements, $\sigma^2_{\text{lp},W}$ is the variance from empirical localization precision in widefield measurements, $\sigma^2_{\text{le},S}$ is the variance from localization errors in scanning measurements, $\sigma^2_{\text{le},W}$ is the variance from localization errors in widefield measurements, and $\sigma^2_{\text{pp}}$ is the variance from placement precision. We determine the values of empirical localization precision from the mean variance of 1600 pitch measurements over a time series of 100 images of the aperture array.

The difference of pitch values between scanning and widefield measurements eliminates $\sigma^2_{\text{pp}}$, isolating the independent terms in $\sigma^2_{\text{pitch},S}$ and $\sigma^2_{\text{pitch},W}$,

$\sigma^2_{\text{pitch},S-W} = \sigma^2_{\text{lp},S} + \sigma^2_{\text{lp},W} + \sigma^2_{\text{le},W} + \sigma^2_{\text{le},S}$, (Eq. S9)

and randomizing the correspondence between the scanning and widefield measurements of pitch causes $\sigma^2_{\text{pp}}$ to be independent between the two measurement methods, giving a variance for the difference between the randomized pitch measurements of

$\left(\sigma^2_{\text{pitch},S-W}\right)_{\text{Random}} = \sigma^2_{\text{lp},S} + \sigma^2_{\text{lp},W} + \sigma^2_{\text{le},W} + \sigma^2_{\text{le},S} + 2\sigma^2_{\text{pp}}$. (Eq. S10)

Subtracting Eq. (S9) from Eq. (S10) isolates $\sigma^2_{\text{pp}}$, providing a measure of placement precision that is free from empirical localization precision and localization error. The corresponding value of placement precision is $\frac{\sigma_{\text{pp}}}{\sqrt{2}}$, where dividing by $\sqrt{2}$ converts pitch standard deviation to position standard deviation. Values for these quantities are in Tables S6 and S7.

Inserting the values of $\sigma^2_{\text{pp}}$ and $\sigma^2_{\text{lp},W}$ into Eq. (S8) gives a localization error in widefield measurements of $\frac{\sigma_{\text{le},W}}{\sqrt{2}}$. Values for these quantities are in Table S7. Subsequent analysis of registration errors indicates that this calculation is conservative, as the localization error evidently includes systematic effects that cancel in registration.

Values from an analogous analysis for scanning measurements of pitch are in Table S6. The widefield values and their components in Table S6 are consistent with but slightly lower than the corresponding values in Table 1 in the main text. This is due to small differences in the characterization of position error by either the ideal array method or measurements of pitch, as well as the exclusion of shot noise.

The measurement uncertainties of variance values are the standard error of the variance as per Ref. [50] in the main text. To determine values of $\sigma_{pp}$, $\sigma_{le,W}$, and $\epsilon_W$, we propagate uncertainty using either the NIST Uncertainty Machine, which is Ref. [52] in the main text, or the law of propagation of uncertainty.

**Table S6.** Pitch variability

| Measurement type | $\sigma^2_{pitch}$ (nm$^2$) | $\sigma^2_{lp}$* (nm$^2$) | $\sigma^2_{le}$ (nm$^2$) |
|---|---|---|---|
| | x direction | | |
| Widefield | 6.83 ± 0.34** | 0.184 ± 0.002** | 0.78 ± 0.50*** |
| Scanning | 7.42 ± 0.37** | 0.138 ± 0.0006** | 1.41 ± 0.52*** |
| | y direction | | |
| I Widefield | 7.73 ± 0.39** | 0.154 ± 0.001** | 1.03 ± 0.54*** |
| Scanning | 7.25 ± 0.36** | 0.131 ± 0.0006** | 0.57 ± 0.52*** |

*Mean variance of 800 values of pitch from a series of 100 images.
**Standard error.
***NIST Uncertainty Machine.

**Table S7.** From pitch variance to position standard deviation

| Quantity | x direction | y direction |
|---|---|---|
| $\sigma^2_{pitch,S-W}$ (nm$^2$) | 2.51 ± 0.13* | 1.88 ± 0.09* |
| $\left(\sigma^2_{pitch,S-W}\right)_{Random}$ (nm$^2$) | 14.25 ± 0.71* | 14.98 ± 0.75* |
| $\sigma^2_{pp}$ (nm$^2$) | 5.87 ± 0.36** | 6.55 ± 0.38** |
| $\sigma_{pp}/\sqrt{2}$ (nm) | 1.71 ± 0.05** | 1.81 ± 0.05** |
| $\sigma_{le,W}/\sqrt{2}$ (random) (nm) | 0.62 ± 0.20*** | 0.72 ± 0.19*** |

*Standard error.
**NIST Uncertainty Machine.
***Propagation of uncertainty.

**Table S8.** Effects of chromatic aberration

| Peak wavelength (nm) | Mean value of image pixel size (nm) | Position of optimal focal plane (nm) |
|---|---|---|
| 400 | 99.85 | 370 |
| 500 | 100.01 | 0 |
| 630 | 100.13 | -720 |

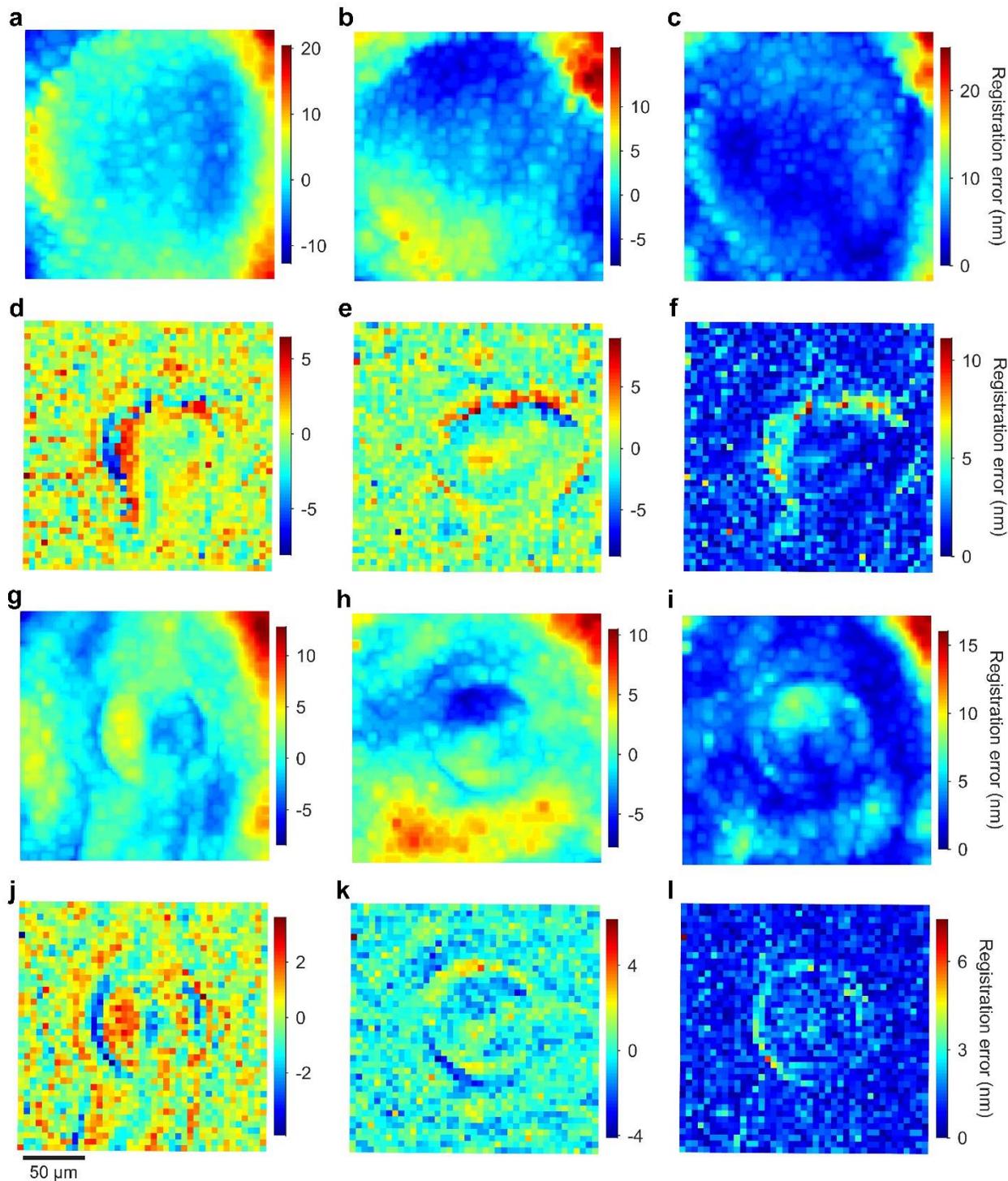

**Figure S15.** Registration errors from three colors at one focal plane. (**a-f**) Plots showing registration errors in (a,d) the x direction, (b,e) the y direction, and (c,f) total magnitude, (a-c) before correction and (d-f) after correction of data from 500 nm and 630 nm peak wavelengths, at the optimal focal plane for the former. (**g-l**) Plots showing registration errors in (g,j) the x direction, (h,k) the y direction, and (i,l) total magnitude (g-i) before correction and (j-l) after correction of data from 400 nm and 500 nm peak wavelengths, at the optimal focal plane for the former.

Systematic errors due to the wavelength dependence of distortion are apparent in the data before correction (a-f, h-j). Systematic errors due to defocus are apparent in the (a-f) 630 nm data and (h-m) 400 nm data.

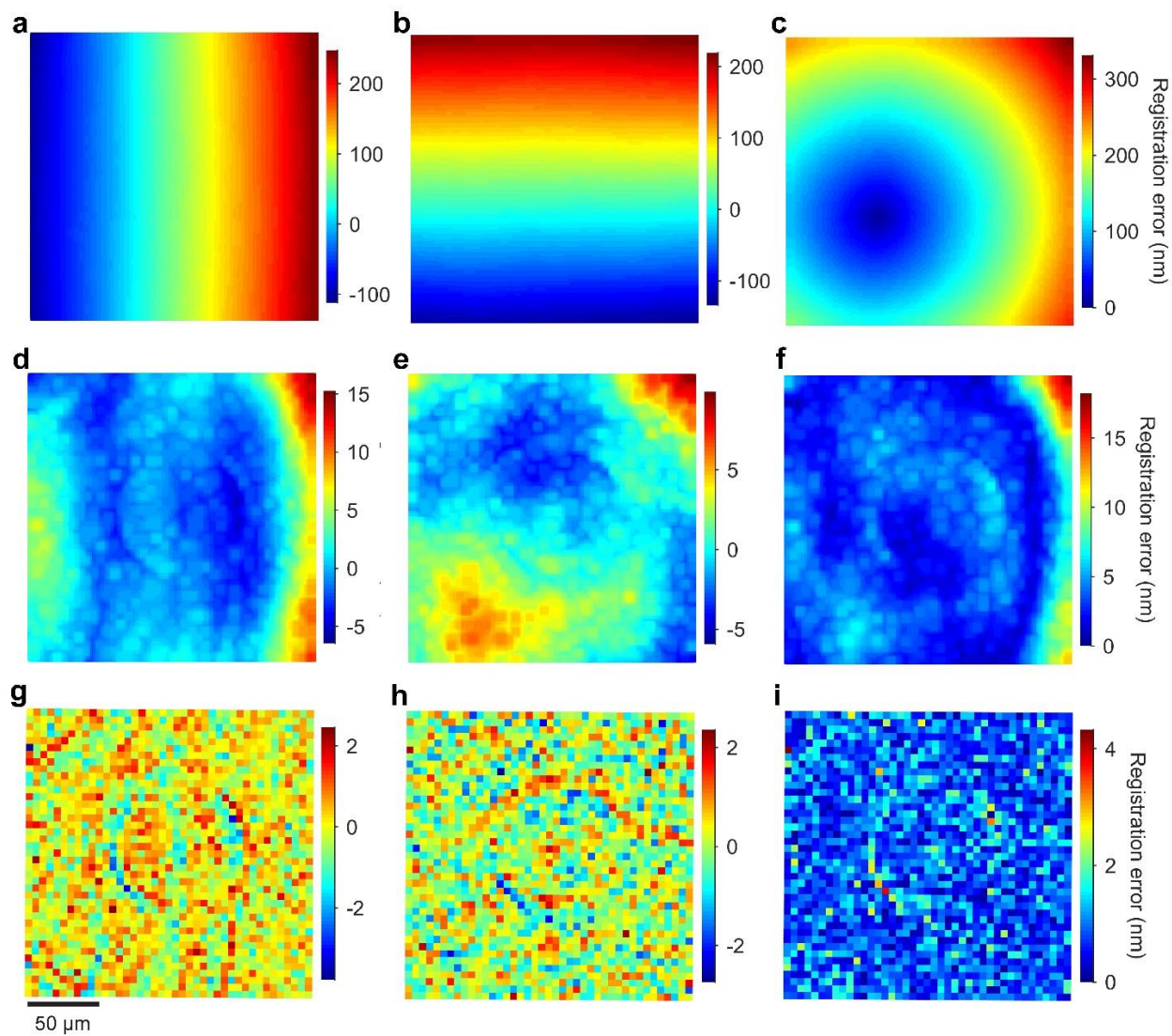

**Figure S16.** Registration errors from two colors at optimal focal planes. **(a-c)** Plots showing registration errors in (a) the x direction, (b) the y direction, and (c) total magnitude, due mostly to different mean values of image pixel size and a lateral offset for localization data from 400 nm and 500 nm peak wavelengths. **(d-f)** Plots showing registration errors in (d) the x direction, (e) the y direction, and (f) total magnitude, after applying a similarity transform to the localization data, due mostly to variable distortion from chromatic aberration. **(g-i)** Plots showing registration errors in (g) the x direction, (h) the y direction, and (i) total magnitude, after applying correction models to the localization data before a similarity transform, due mostly to localization error and empirical localization precision.
46

**Note S9.** Error analysis for multicolor registration

Registration errors of data after correction from two colors are due to a combination of empirical localization precision and localization error, having a variance of

$$\sigma_{reg}^2 = \sigma_{lp,1}^2 + \sigma_{lp,2}^2 + \sigma_{le,1}^2 + \sigma_{le,2}^2, \quad \text{(Eq. S12)}$$

where $\sigma_{lp,1}^2$ and $\sigma_{lp,2}^2$ are the variance due to empirical localization precision, and $\sigma_{le,1}^2$ and $\sigma_{le,2}^2$ are the variance due to localization error for colors 1 and 2, respectively. Assuming the localization error is the same for each color channel, or equivalently considering the mean value, and by measuring the empirical localization precision, we determine the contribution of localization error to the registration error as

$$\sigma_{le} = \sqrt{\frac{\sigma_{reg}^2 - \sigma_{lp,1}^2 - \sigma_{lp,2}^2}{2}}. \quad \text{(Eq. S13)}$$

Values of empirical localization precision are in Table S9. Values of the contribution of localization error to registration error, $\sigma_{le}$, for data before and after correction prior to registration are in Table S10.

**Table S9.** Empirical localization precision in multicolor registration

| Peak wavelength (nm) | $\sigma_{lp,x}$ (nm) | $\sigma_{lp,y}$ (nm) |
|---|---|---|
| 400 | 0.340 ± 0.003 | 0.318 ± 0.002 |
| 500 | 0.371 ± 0.003 | 0.315 ± 0.002 |
| 630 | 0.394 ± 0.002 | 0.320 ± 0.002 |

Uncertainties are one standard error.

**Table S10.** Localization error in multicolor registration

| | | 400 nm and 500 nm | | 500 nm and 630 nm | |
|---|---|---|---|---|---|
| | | $\sigma_{le,x}$ (nm) | $\sigma_{le,y}$ (nm) | $\sigma_{le,x}$ (nm) | $\sigma_{le,y}$ (nm) |
| Optimal focal planes | Uncorrected | 2.23 ± 0.04 | 1.70 ± 0.03 | 2.45 ± 0.04 | 1.78 ± 0.03 |
| | Corrected | 0.40 ± 0.01 | 0.41 ± 0.01 | 0.35 ± 0.01 | 0.47 ± 0.01 |
| Single focal plane | Uncorrected | 1.85 ± 0.03 | 1.85 ± 0.03 | 2.86 ± 0.05 | 2.86 ± 0.05 |
| | Corrected | 0.63 ± 0.01 | 0.59 ± 0.01 | 1.16 ± 0.02 | 1.28 ± 0.02 |

Uncertainties are one standard error.

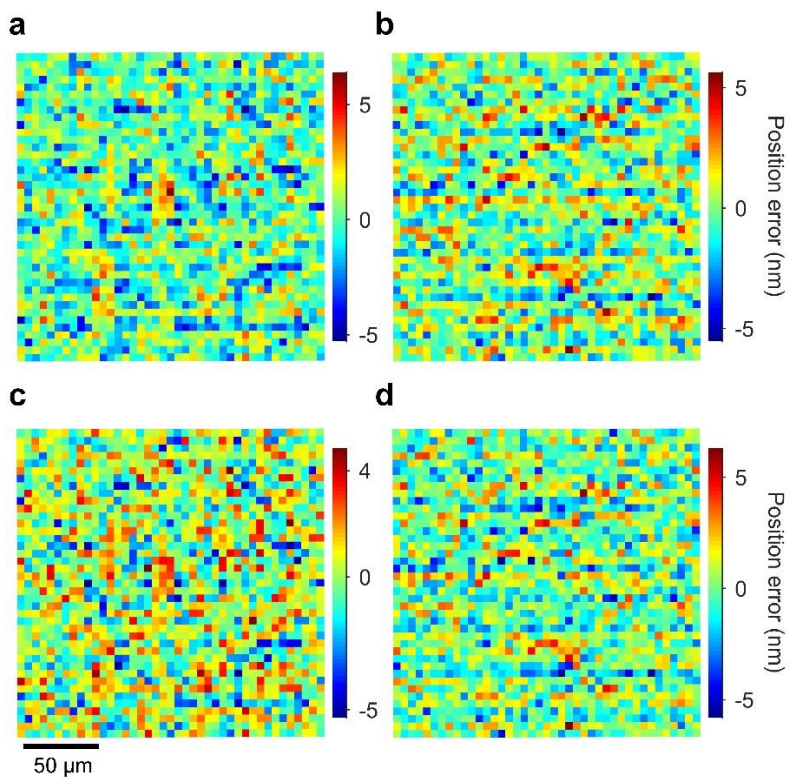

**Figure S17.** Correction of fluorescence data. (**a-d**) Plots showing position errors in (a,c) the x direction and (b,d) the y direction following correction of data from (a-b) transillumination and (c-d) fluorescence. These results show that our reference materials and calibration methods are equally applicable to transillumination of empty apertures and epiillumination of fluorescent dye in apertures.

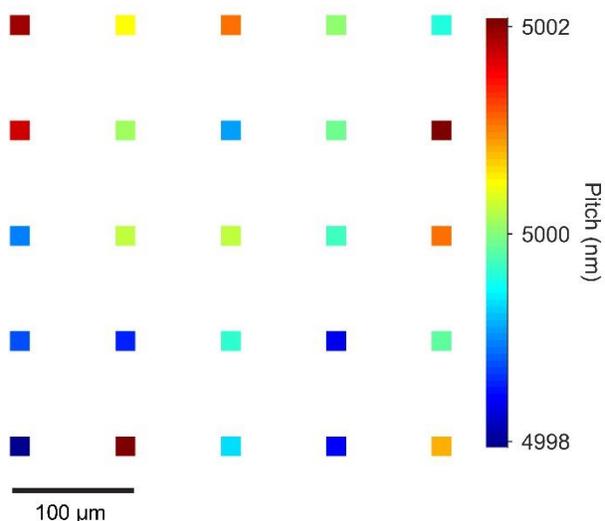

**Figure S18.** Pitch across the aperture array. Plot showing 25 regions of the aperture array, with color scale indicating the mean pitch from four aperture pairs within each region. Data marker size is not to scale. No systematic effects indicative of electron-optical aberrations are evident.

**Table S11.** Pitch characterization for two lithography systems

|  | x direction | | y direction | |
|---|---|---|---|---|
|  | **Array 1** | **Array 2** | **Array 1** | **Array 2** |
| Mean pitch (pixels) | 49.969 | 49.958 | 49.974 | 49.964 |
| Standard error (pixels) | 0.003 | 0.002 | 0.002 | 0.002 |

**Note S10.** Rigidity analysis

The positions of each aperture or nanoparticle define a nominally rigid constellation of points in the image plane, $(x_{j,\eta}, y_{j,\eta})$, where the index $j$ denotes an image in a measurement series and the index $\eta$ denotes a point in a constellation[5]. We measure and remove the common-mode motion of the sample by applying a two-dimensional rigid transformation to map the constellation in image $j$ to the constellation in image $k$. This transformation consists of a displacement of the centroid of the constellation $(X_j - X_k)\hat{x} + (Y_j - Y_k)\hat{y}$ and a rotation of the constellation about the centroid, $\Delta\theta = \theta_j - \theta_k$, where $(X_j, Y_j)$ and $(X_k, Y_k)$ are the positions of the centroids in images $j$ and $k$, respectively, and $\theta_j$ and $\theta_k$ are the orientations of the constellation in images $j$ and $k$, respectively. The optimal rigid transformation minimizes the registration error between corresponding points in images $j$ and $k$. Registration error is insensitive to systematic errors in localizing single apertures or nanoparticles. Therefore, we omit CMOS calibration from this analysis.

Motion of a sample in the z direction during a time series can cause apparent deformation of a rigid constellation in optical micrographs. At time scales that allow, we minimize these effects by imaging through focus at each point in the time series, acquiring images at multiple z positions around the plane of optimal focus for the entire time series. The nominal spacing in z position between each image is 10 nm, set by the resolution of our piezoelectric nosepiece that controls the position of the objective lens. At each time point, we choose from the set of images at varying z positions the one image that minimizes the root-mean-square of the registration errors from registration with the first image in the time series. This procedure minimizes any motion of the sample in the z direction relative to the position at the initial time point, so that the images that form the resulting time series share a common z position within 10 nm.

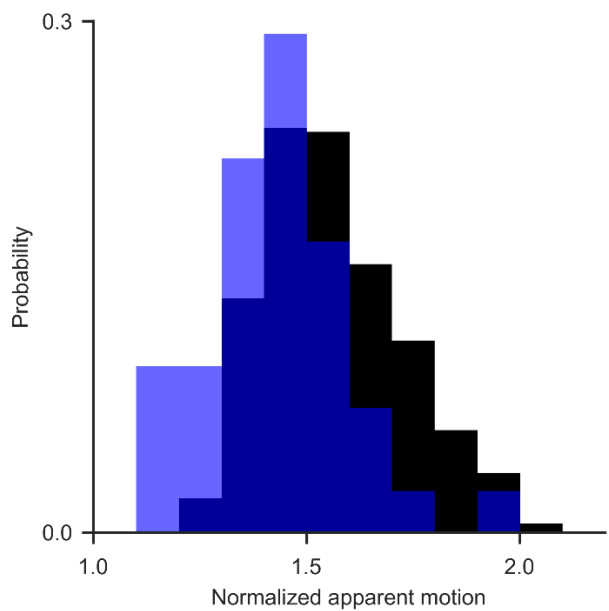

**Figure S19.** Nanoparticle stability down to $10^{-1}$ s. Plot showing probability distributions of normalized apparent motion for nominally motionless apertures (black) and nanoparticles (blue) that we image at a frequency of $10^1$ s$^{-1}$ for a duration of $10^1$ s, without intentionally changing the z position. The normalization is with respect to the Cramér–Rao lower bound and accounts primarily for differences in the number of signal photons. The corresponding absolute mean values define the measurement uncertainties, and are approximately 0.43 nm for apertures and 0.55 nm for nanoparticles. The magnitude of normalized apparent motion for nanoparticles is comparable to that of static apertures, indicating that the nanoparticles are also static at these scales.

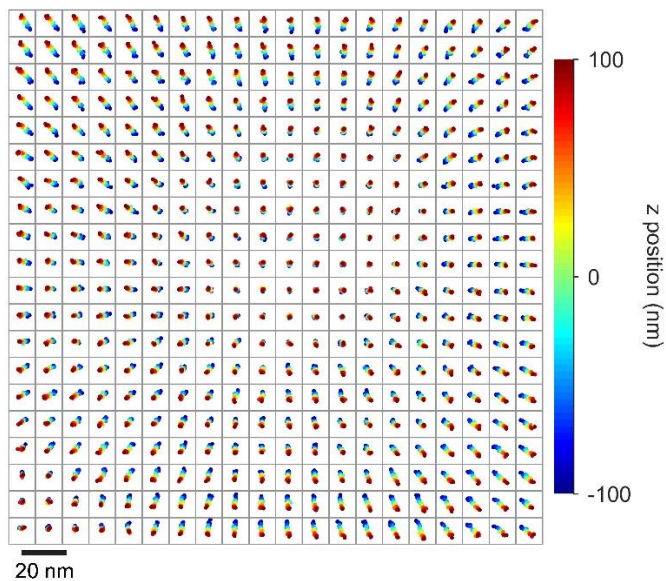

**Figure S20.** Apparent motion. Grid of scatterplots, each corresponding to a single aperture, showing apparent motion in the radial direction due to imaging through focus over a range of 200 nm in z position. The grid spacing indicates an array pitch of 10 µm. The scale bar corresponds to the scatterplots.

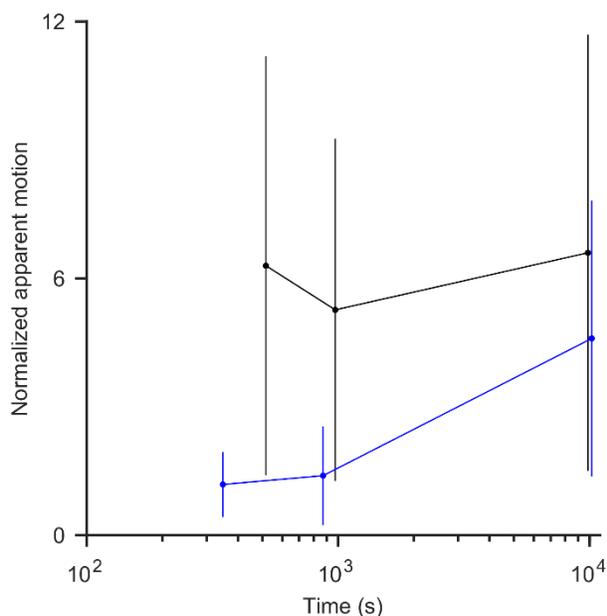

**Figure S21.** Nanoparticle stability up to $10^4$ s. Plot showing normalized apparent motion as a function of time, exceeding the time that is necessary for imaging through focus, for nominally static apertures (black) and nanoparticles (blue). Normalization is with respect to empirical localization precision, or the corresponding values of apparent motion at the time scale of $10^{-1}$ s. Data markers are mean values and vertical bars are ± one standard deviation. The values of normalized apparent motion for nanoparticles are comparable to those of apertures, indicating that the nanoparticles are static at these scales.